\documentstyle[prl,aps,epsfig,amssymb,twocolumn]{revtex}

\begin{document}

\newcommand{\beq}{\begin{equation}}
\newcommand{\eeq}{\end{equation}}
\newcommand{\barr}{\begin{eqnarray}}
\newcommand{\earr}{\end{eqnarray}}

\newcommand{\andy}[1]{ }

\newcommand{\bm}[1]{\mbox{\boldmath $#1$}}
\newcommand{\bmsub}[1]{\mbox{\boldmath\scriptsize $#1$}}

\def\bra#1{\langle #1 |}
\def\ket#1{| #1 \rangle}
\def\sinc{{\rm sinc}}

\title{Mesoscopic interference}

\author{ P. FACCHI, A. MARIANO and S. PASCAZIO}

\address{Dipartimento di Fisica, Universit\`a di Bari
and Istituto Nazionale di Fisica Nucleare, Sezione di Bari \\
 I-70126  Bari, Italy }

\date{\today}

\maketitle

\draft

\begin{abstract}
We analyze a double-slit experiment when the interfering particle
is ``mesoscopic" and one endeavors to obtain {\em Welcher Weg}
information by shining light on it. We derive a compact expression
for the visibility of the interference pattern: coherence depends
on both the spatial and temporal features of the wave function
during its travel to the screen. We set a bound on the temperature
of the mesoscopic particle in order that its quantum mechanical
coherence be maintained.
\end{abstract}


\setcounter{equation}{0}
\section{Introduction}
\label{sec-introd}
\andy{intro}
The double-slit experiment is one of the simplest and most
fundamental examples in quantum mechanics. However, in spite of
its simplicity, its explanation is subtle and brings to light some
of the most intriguing features of the quantal description of
nature. According to Feynman, Leighton and Sands \cite{Feynman}
``[this is] a phenomenon which is impossible, {\em absolutely}
impossible, to explain in any classical way, and which has in it
the heart of quantum mechanics. In reality, it contains the {\em
only} mystery."

It is not exaggerated to say that our comprehension of nature has
been shaped by our advances in interference and interferometry,
both at an experimental and a theoretical level. Technological
progress has played a primary role: experiments that were
undreamed of until a few years ago can be carried out nowadays.
Double-slit interference experiments with photons
\cite{MandelWolf}, neutrons \cite{RauchWerner}, electrons
\cite{Tonomura}, atoms \cite{atoms} and small molecules
\cite{molecules,molecules1} can now be routinely performed. All
these systems can be considered microscopic, essentially because
they are ``elementary," can be described in terms of a wave
function and their evolution is governed by the Schr\"odinger
equation with amazing accuracy.

The aim of the present paper is to discuss the interference of
{\em mesoscopic} systems. ``Mesoscopic" objects are neither {\em
micro}scopic nor {\em macro}scopic. Although we shall not attempt
to give a rigorous definition of ``mesoscopicity" (we do not know
any), we shall think of systems that can be described by a wave
function, yet are made up of a significant number of elementary
constituents, such as atoms. Most important, they are
characterized by a nontrivial internal structure that can have
both quantal and classical features. A significant example, on
which we shall focus our attention, is a molecule of fullerene,
made up of 60 nuclei of Carbon and 360 electrons, for a total of
about $10^3$ ``elementary" constituents. Although fullerenes are
fully quantum mechanical systems, they also have macroscopic-like
features and emit thermal (blackbody) radiation
\cite{C60e,C60,Kolodney}. Very recently the quantum interference
of fullerene molecules (C$_{60}$ and C$_{70}$) has been observed
in a series of pioneering experiments performed in Vienna
\cite{AZfull,AZcomptes}. Our aim is to analyze the interference of
fullerene from a theoretical and fully quantum mechanical
viewpoint.

The discussion that follows is of general validity. However, as
S.J.\ Gould masterly suggests \cite{Gould} ``frontal attacks upon
generalities inevitably lapse into tedium or tendentiousness. The
beauty of nature lies in detail; the message, in generality.
Optimal appreciation demands both, and I know no better tactic
than the illustration of exciting principles by well-chosen
particulars." Although our analysis applies to any molecule or
system endowed with an internal structure, the kind of questions
that pop up in one's mind when one ponders over the properties and
complexity of C$_{60}$ make the following discussion very
fascinating.

This paper contains tutorial sections as well as original
material. We start by setting up the notation and outlining the
physics of the double-slit experiment in Section 2. Although the
content of this section is usually the subject of elementary
textbooks of quantum mechanics, our analysis is original: we
derive the double-slit diffraction pattern by solving the
time-dependent Schr\"odinger equation, with given (and well
chosen) initial conditions. In Section 3 we revisit Heisenberg's
microscope \cite{Heisenberg}, analyzing an experimental
configuration aimed at determining which slit the particle goes
through. This example as well can be found in introductory
chapters of textbooks of quantum mechanics; however, we introduce
here a novel element of discussion, ``postponing" the
determination of the particle's route in a way that will turn out
to be interesting and significant for the subsequent analysis. The
central part of the manuscript are Sections 4 and 5. Section 4
contains a model calculation leading to an (almost) exact formula
for the visibility of the interference pattern when a {\em
complex} molecule (``fullerene") passes through a double slit and
is illuminated by laser light of given wavelength. The visibility
of the interference pattern depends on the laser wavelength (as
expected), but also, interestingly, on the lifetime of the
reemission process. We tried to keep the discussion at a
reasonably elementary level, adding two appendices in which some
relevant notions of advanced quantum mechanics and quantum
electrodynamics are explained. Section 5 contains a discussion of
what we will call ``interference of mesoscopic systems." A novel
formula will be derived, relating the features of the interference
pattern to the temperature of the fullerene molecule. A
``decoherence temperature" will be defined as a function of some
intrinsic properties of the molecule (such as its area and
velocity) and the geometry of the experimental setup (such as the
separation of the slits and the distance between slits and
screen). Section 6 is devoted to our concluding remarks. Our
analysis will motivate us to ask in which sense a molecule of
fullerene can be considered ``microscopic" and, by contrast, when
it is more appropriate to think of it in mesoscopic terms. This
will lead us to wonder about the significance of quantum
(de)coherence.

\section{Double-slit interference}
\label{sec-2slits}
\andy{2slits}

\begin{figure}
\begin{center}
\epsfig{file=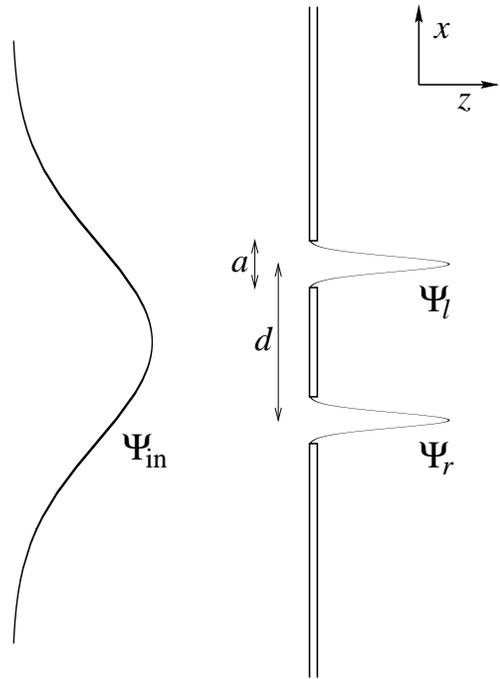,width=6.5cm}
\end{center}
\caption{A quantum object impinging on a double slit.
The velocity of the wave packet is in direction $+z$, the slits
have width $a$ and are separated by a distance $d$.}
\label{fig:gauss2sl}
\end{figure}

We start by looking at the simplest quantum mechanical experiment:
consider a quantum system described by a wave packet $\Psi_{\rm
in}$, impinging on a double slit. We assume that the wave packet
travels along direction $+z$ and its transverse coherence length
is larger than the distance between the slits, so that the two
wave packets emerging from the slits are coherent with each other.
This is a fundamental requirement in interferometry, both at a
quantal and classical level (although the quantal situation has a
different ``charm" if one thinks that the experiment is performed
by accumulating events pertaining to single quantum systems
\cite{Dirac}). The slits are parallel to $y$, have width $a$ and
are separated by a distance $d$, along direction $x$. The geometry
of our arrangement is outlined in Figure \ref{fig:gauss2sl}.

We shall assume that the wave functions (the one impinging on the
slit and those emerging from it) can be approximated by Gaussians.
In general, a Gaussian wave packet has the form
\andy{gaussgen}
\beq
\bra{\bm x}\Psi\rangle=\Psi(\bm x,t=0)=\psi_x(x)\psi_y(y)\psi_z(z)
\label{eq:gaussgen}
\eeq
where $\ket{\Psi}$ is the quantum state and
\andy{gaussgenxyz}
\barr
\psi_x(x)\!\! &=& \!\!\frac{1}{(2\pi\delta x^2)^{1/4}} \nonumber \\
& & \times \exp
\left(-\frac{1-i\eta_x}{4\delta x^2}(x-\bar x)^2
+\frac{i}{\hbar}\bar p_x x-i\phi_x\right) \nonumber \\
\label{eq:gaussgenxyz}
\earr
(analogously for $y$ and $z$), where $\eta_x$ and $\phi_x$ are real
constants, $\bar x$ and $\bar p_x$ are the average position and
momentum, respectively, and their standard deviations $\delta x$ and
$\delta p_x$ satisfy the relation
\andy{mindefect}
\beq
\delta x
\delta p_x=(1+\eta_x^2)^{\frac{1}{2}} \frac{\hbar}{2}\geq \frac{\hbar}{2},
\label{eq:mindefect}
\eeq
so that (\ref{eq:gaussgenxyz}) has the minimum uncertainty for
$\eta_x=0$. The evolution of the packet (\ref{eq:gaussgen}) in
free space $\bra{\bm x}e^{-ip^2t/2m\hbar}\ket{\Psi}$ is readily
evaluated and maintains a Gaussian form. One gets
\andy{evgaussgen}
\barr
\psi_x(x,t)&=&
\frac{1}{(2\pi\delta x^2(t))^{1/4}} \nonumber \\
& & \times \exp
\left(-\frac{1-i\eta_x(t)}{4\delta x^2(t)}(x-\bar x(t))^2
\right. \nonumber \\
& &
\left. + \ \frac{i}{\hbar}\bar p_x x-i\phi_x(t)\right),
\label{eq:evgaussgen}
\earr
where
\andy{evolmean}
\barr
\bar x(t)&=&\bar x(0)+\frac{\bar p_x}{m}t,\nonumber\\
\eta_x(t)&=&\eta_x(0)+2\frac{\delta p_x^2}{m\hbar} t,
\nonumber \\
\delta x^2(t)&=&\frac{\hbar^2}{4\delta p_x^2}(1+\eta_x^2(t)),
\label{eq:evolmean}\\
\phi_x(t)&=&\phi_x(0)+\frac{\bar p_x^2}{2 m\hbar} t
\nonumber \\
 & & + \ \frac{1}{2}\left(\arctan\eta_x(t)-\arctan\eta_x(0)\right).
\nonumber
\earr
Obviously, the spread $\delta p_x$ and average momentum $\bar p_x$
remain unchanged during the free evolution.

The preceding formulas are of general validity. Our initial state
is that emerging from the two slits
\andy{gausslr}
\beq
|\Psi_0\rangle = \frac{1}{\sqrt{N}}(
|\Psi_\ell\rangle+|\Psi_r\rangle),
\label{eq:gausslr}
\eeq
where $\ell$ and $r$ stand for ``left" and ``right" respectively,
$N$ is a normalization factor (see later) and
\andy{gauss}
\beq
\Psi_{\ell,r}(\bm x)=
\langle \bm x|\Psi_{\ell,r}\rangle = \psi_{\ell,r}(x) \psi_y(y)
\psi_z(z).
\label{eq:gauss}
\eeq
All wave functions have the Gaussian form (\ref{eq:gaussgenxyz})
with
\andy{parm}
\barr
\bar x_{\ell,r}(0)&=&\mp\frac{d}{2}, \nonumber \\
\bar y(0) &=& \bar z(0)=0, \nonumber \\
\bar p_z&=&\hbar k_0, \label{eq:parm} \\
\bar p_x &=& \bar p_y=0, \nonumber \\
N &=&\parallel \ket{\Psi_\ell}+\ket{\Psi_r} \parallel^2 \nonumber
\\ & = & 2\left(1+\exp\left(-\frac{d^2\delta
p_x^2}{2\hbar^2}\right)\right) \nonumber ,
\earr
so that the initial average positions of the two wave packets are
$(\mp d/2,0,0)$ and their average momentum $\hbar k_0$ is in
direction $+z$. We see that $N\simeq 2$ if the left and right
packets are well separated ($d\gg\hbar/\delta p_x$). As already
stressed, we are assuming that the wave function after the slits
can be written as a ``double Gaussian." This is, for example, the
approach of Feynman and Hibbs \cite{FeynmanHibbs}. The evolution
yields
\andy{evpsi}
\beq
\ket{\Psi(t)}= e^{-ip^2t/2m\hbar}\ket{\Psi_0}
\label{eq:evpsi}
\eeq
and interference is observed at a screen perpendicular to $z$,
placed at a distance $z=L$ from the plane of the slits. The
problem becomes essentially one dimensional (the relevant
coordinate being $x$) and the position probability distribution at
the screen reads
\andy{gaussscr}
\barr
& & \!\!\!\!\!\! |\Psi(\bm x,t_0)|^2 \nonumber \\
& &  = |\bra{\bm x} \Psi (t_0)\rangle|^2 \nonumber \\
& & \equiv \frac{1}{N}|\Psi_{\ell}(\bm x,t_0)+\Psi_{r}(\bm
x,t_0)|^2
\nonumber \\
& & = \frac{1}{N\sqrt{2\pi\delta x^2(t_0)}}
\left[
\exp\left(-\frac{\left(x+\frac{d}{2}\right)^2}{2\delta x^2(t_0)}\right) \right.
\nonumber \\
 & & \left. \quad + \
\exp\left(-\frac{\left(x-\frac{d}{2}\right)^2}{2\delta x^2(t_0)}\right)
\right. \nonumber \\ & &
\left. \quad + \ 2 \exp\left(-\frac{x^2+\left(\frac{d}{2}\right)^2}{2\delta x^2(t_0)}\right)
\cos \left(\frac{\eta_x(t_0) d}{2\delta x^2(t_0)}
\;x\right)
\right]
\nonumber \\
 & &
\quad \times \; |\psi_y(y,t_0)|^2|\psi_z(z,t_0)|^2,\nonumber\\
\label{eq:gaussscr}
\earr
where $t_0 =mL/\hbar k_0$ is the time of arrival of the wave packet
at the screen.

This analysis is of general validity. However, in order to
concentrate our attention on a concrete physical problem, we shall
focus on the experiment \cite{AZfull} and take the slits to have
width $a$ and to be separated by a distance $d=2a$. The intensity
at the screen is
\andy{intensity}
\barr
I(x)&=&\bra{\Psi(t_0)} x\rangle\bra{x}\Psi(t_0)\rangle
\nonumber \\
&=& \int dy dz |\Psi(x,y,z,t_0)|^2,
\label{eq:intensity}
\earr
which, due to normalization, is equal to (\ref{eq:gaussscr})
without the factor $|\psi_y(y,t)|^2|\psi_z(z,t)|^2$. We set
$a=50$nm, $d=100$nm, $L=1.22$m, $m=1.197\cdot 10^{-24}$kg and
consider a beam with $\bar v_z=128$m/s, so that one gets $k_0=\bar
v_z m/\hbar=1.46\cdot10^{12}$m$^{-1}$, $\lambda_0=4.3$pm and
$t_0=9.47$ms. These values are taken from the latest Vienna
experiment \cite{AZcomptes}. One of the advantages of focusing on
a concrete physical example is that one gets a feeling for the
numbers. This is particularly important when one deals with
systems that can be properly considered mesoscopic (fullerene has
a mass $m\simeq 720$u and is made up of $\simeq 10^3$ particles).
One of the main ideas to be discussed in this paper is that
coherence (i.e.\ the possibility of observing an interference
pattern) is a {\em quantitative} issue: if the experimenter is
able to keep under control all disturbances/noises/interactions in
the setup, even a mesoscopic (or, in principle, macroscopic)
system will preserve its coherence and display a double-slit
interference pattern. We shall come back to this point in the next
sections.

\begin{figure}
\begin{center}
\epsfig{file=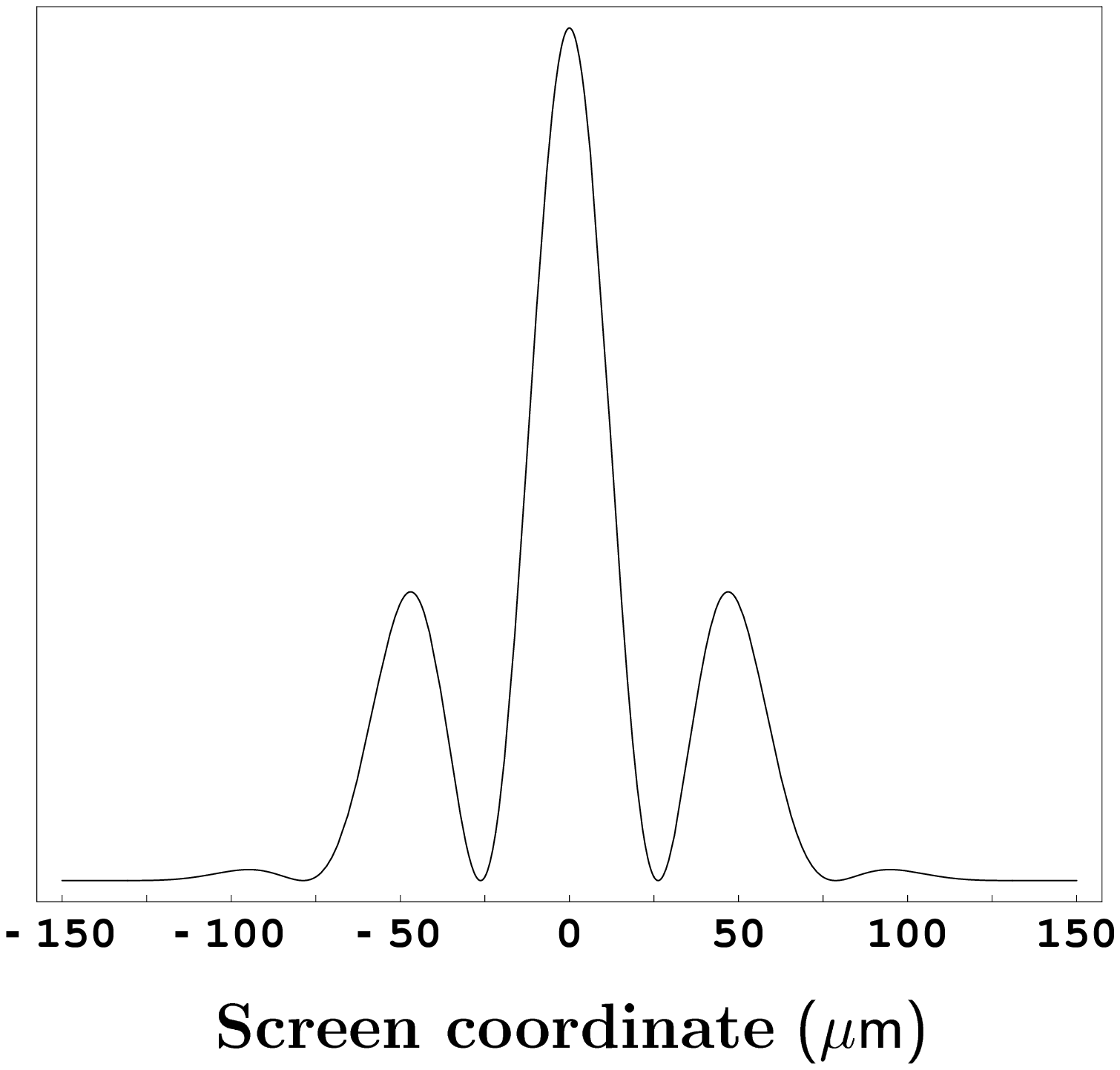,width=6.45cm}
\end{center}
\caption{The interference pattern (\ref{eq:freepatt}).}
\label{fig:pattern}
\end{figure}

In the case considered above, we can choose $\delta x(0)\simeq a$.
The results one obtains are completely independent of this choice
if one looks at the far-field interference pattern, namely $\delta
x(t_0)\gg\delta x(0)$, i.e. $\eta_x(t_0)\gg\eta_x(0)$. This is our
case and we get from Eqs.\ (\ref{eq:evolmean}) (far field)
\andy{etadelta}
\barr \eta_x(t_0) & \simeq & 2\frac{\delta p_x^2}{m\hbar} t_0,
\nonumber \\
\delta x(t_0) & \simeq & \frac{\hbar}{2}\frac{\eta_x(t_0)}{\delta p_x}
\simeq \frac{\delta p_x}{m} t_0 \label{eq:etadelta}
\earr
and
\andy{capX}
\barr
& & \frac{\eta_x(t_0) d}{2\delta x^2(t_0)}
\simeq\frac{d m}{\hbar t_0}
=\frac{2\pi}{X}, \label{eq:capX} \\
& & \quad\mbox{with}\quad X=\frac{h t_0}{m d}=\frac{2 \pi L}{k_0
d} = 52.46\mu {\rm m} .
\nonumber
\earr
Hence we can rewrite the intensity pattern
\andy{freepatt}
\beq
I(x)\simeq\frac{e^{-x^2/2\delta x^2(t_0)}}{\sqrt{2\pi\delta
x^2(t_0)}}
\left[1+\cos\left(2\pi\frac{x}{X}\right)\right],
\label{eq:freepatt}
\eeq
where we neglected $a=d/2$ with respect to $\delta x(t_0)$ in the
Gaussian envelope functions in (\ref{eq:gaussscr}). The intensity
at the screen is shown in Figure \ref{fig:pattern}. Notice that
the interference pattern has been obtained by simply solving the
Schr\"odinger equation (free evolution in vacuum), as it should.
The only free parameter is $\delta p_x$, which is determined by
imposing the dispersion at the screen $\delta x(t_0)=33.7\mu$m,
which implies, by (\ref{eq:etadelta}), $\delta p_x=4.26\cdot
10^{-27}$kg m/s.

It is interesting to observe that the minimum spatial width at the
slits $\delta x(0)$ that can be chosen is constrained by the
uncertainty relations (\ref{eq:mindefect}) with $\eta_x(0)=0$ and
reads $\delta x(0)=12.4\mbox{nm}\simeq a/4$, so that the position
probability density at the slit boundary is reduced to 10\% of its
maximum value. This suggests that the choice of Gaussian wave
packets at the slits is not optimal; a better wave function could
be a bell-shaped function flattened at the top. Notice also that,
as already stressed, the parameter $\delta_x(0)$ does not enter in
the expression of the intensity pattern (\ref{eq:freepatt}) and
only guarantees the internal consistence of the calculation. In
the same spirit of other calculations aimed at analyzing
wave-packet effects
\cite{FeynmanHibbs,Englert,Bohm}, our analysis complements those
based on the plane-wave approximation \cite{BornWolf,Zeilinger}.

\begin{figure}
\begin{center}
\epsfig{file=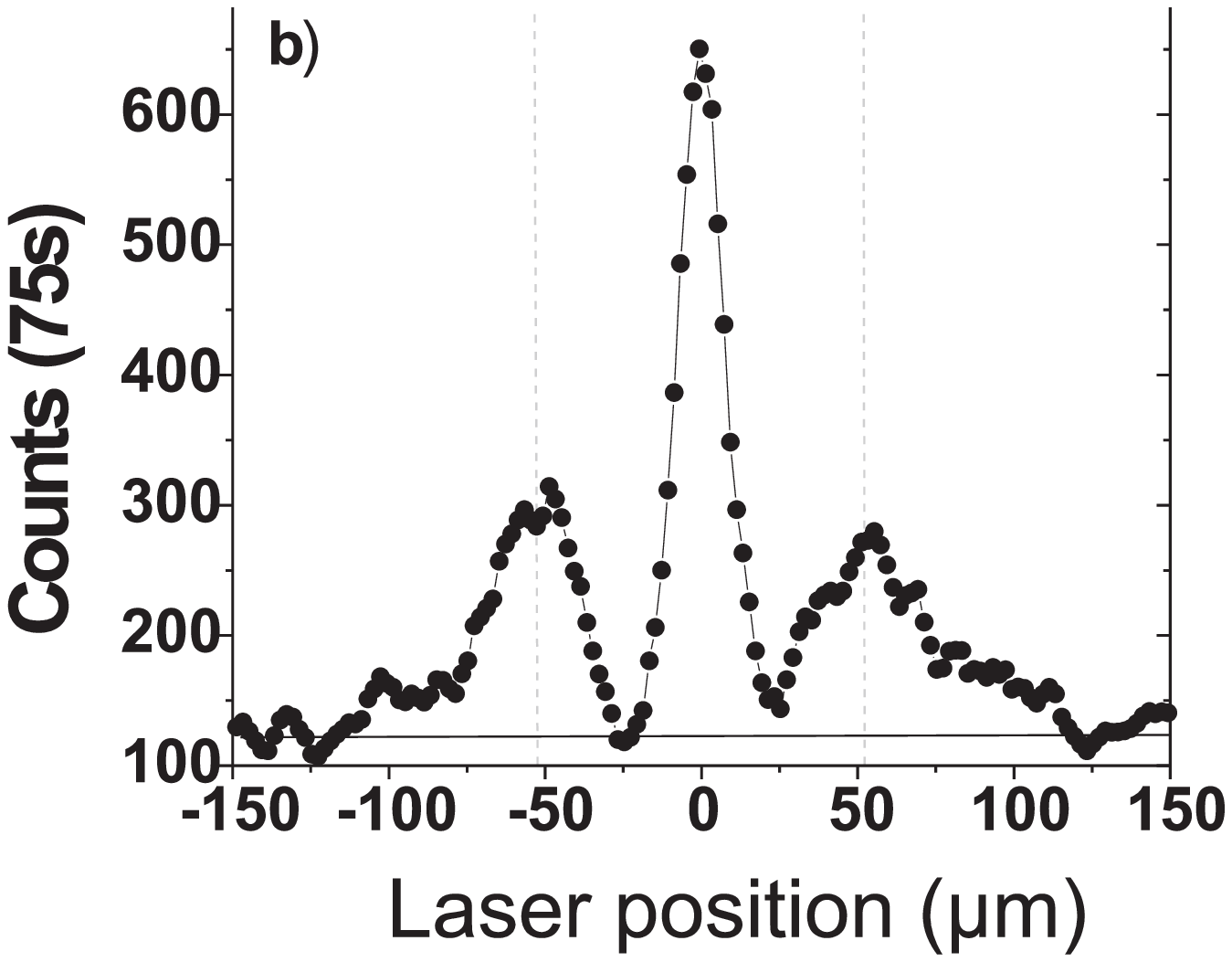,width=8cm}
\end{center}
\caption{Interference of C$_{60}$: experimental results \protect\cite{AZcomptes}.
(Courtesy of the Vienna group.)  Compare with Figure
\ref{fig:pattern}, where the (only) free parameter $\delta x(t_0)$
has been adjusted in order to reproduce the experimental data.
(``Laser position" in this figure is simply the screen
coordinate.)}
\label{fig:exptresult}
\end{figure}

The beautiful experimental results obtained by the Vienna group
are shown in Figure \ref{fig:exptresult}. Notice the high
visibility, obtained with a well collimated molecular beam and a
careful technique of velocity selection \cite{AZcomptes}. The
asymmetry of the data may be ascribed to the velocity selection
technique. By comparing Figure \ref{fig:exptresult} with Figure
\ref{fig:pattern}, obtained in the hypothesis of a double-Gaussian
initial state, by setting the (only) free parameter $\delta
x(t_0)= 33.7\mu$m, one is led to think that only a few (say 2 or
3) slits of the diffraction grating are coherently illuminated by
each fullerene molecule in the beam. The detailed features of a
double-slit experiment for large molecules are still under
investigation and there are interesting proposals concerning a
reduced ``effective" slit width \cite{Hegerfeldt,AZfull}. In our
``minimal" calculation these additional effects have not been
considered.

\section{The Heisenberg-Bohm microscope revisited}
\label{sec-microscopio}
\andy{microscopio}

\begin{figure}[t]
\begin{center}
\epsfig{file=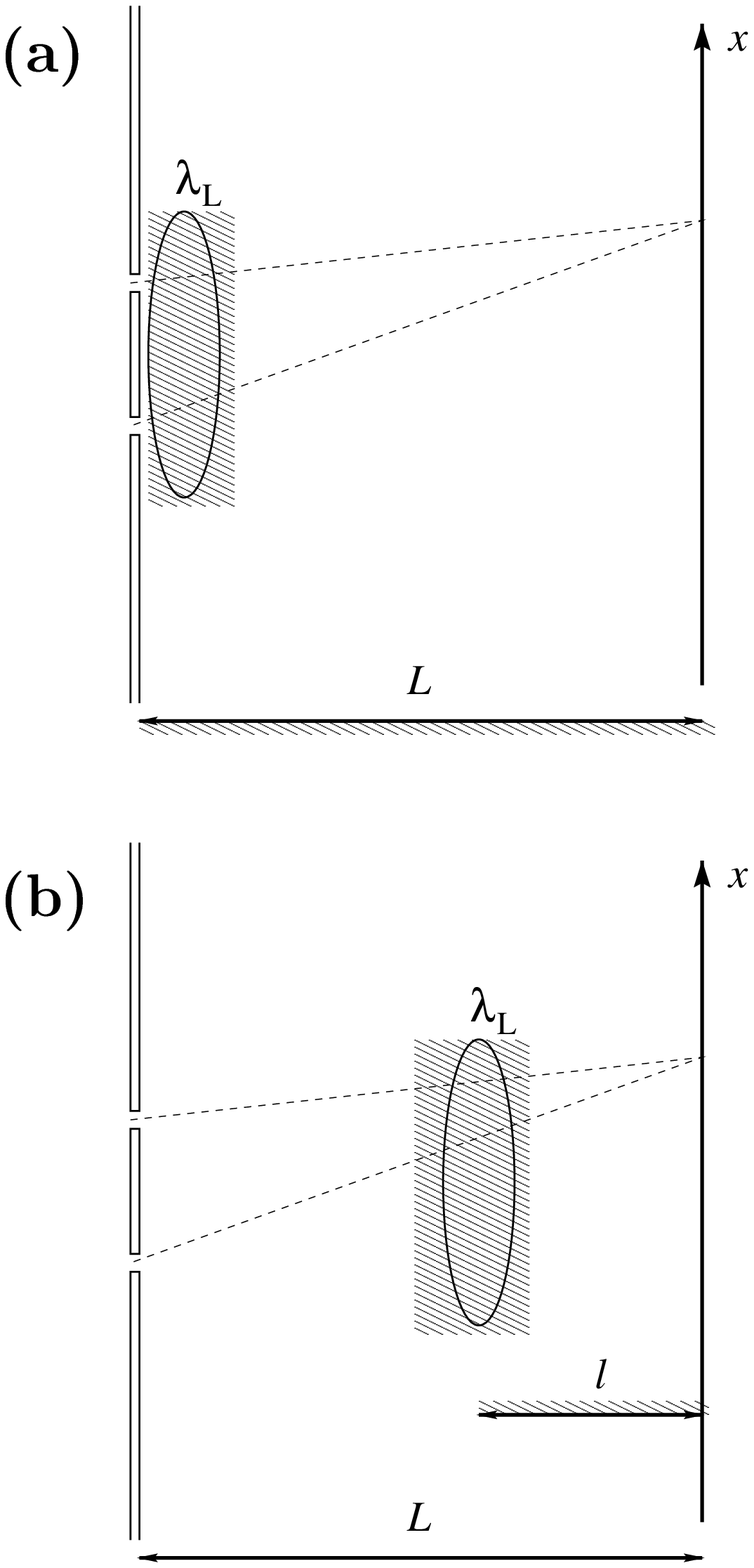,width=5.5cm}
\end{center}
\caption{The Heisenberg-Bohm {\em Welcher Weg} experiment.
(a) Standard setup. (b) Modified version: the laser light is
shined at a distance $L-\ell$ from the slits. }
\label{fig:2sllaser}
\end{figure}

Interference disappears if one endeavors to obtain {\em Welcher
Weg} information. Let us follow Bohm's discussion \cite{Bohm} of a
double-slit version of Heisenberg's microscope \cite{Heisenberg}.
The experiment is sketched in Figure \ref{fig:2sllaser}(a). The
situation is analogous to that described in the preceding section,
but now a laser beam parallel to the slits ($y$ direction) is
shined at the exit of the slits. The laser light has wavelength
$\lambda_{\rm L}$ and the laser spot is larger than $d$, the
distance between the slits. If a photon is scattered off the
interfering particle, the momentum of the latter becomes uncertain
of the quantity
\andy{deltap}
\beq
\triangle p \simeq h/\lambda_{\rm L},
\label{eq:deltap}
\eeq
which ``shakes" the interference pattern at the screen by the
quantity
\andy{deltatheta}
\beq
\triangle \theta \simeq \frac{\triangle p}{p} \simeq
\frac{h}{\lambda_{\rm L}p} \quad \Leftrightarrow
\quad \triangle x \simeq \frac{h}{\lambda_{\rm L}p} L,
\label{eq:deltatheta}
\eeq
where $\theta \simeq x/L$. On the other hand, from Eq.\
(\ref{eq:freepatt}),
\andy{freepattbis}
\beq
I(x)\propto 1+\cos\left(2\pi\frac{x}{X}\right)=
1+\cos\left(\frac{pd \theta}{\hbar}\right) ,
\label{eq:freepattbis}
\eeq
so that the distance between a minimum and the adjacent maximum at
the screen is
\andy{deltathetan}
\beq
\triangle \theta_M = \frac{h}{2pd} \quad \Leftrightarrow \quad
\triangle x_M \simeq \frac{h}{2pd} L
\label{eq:deltathetan}
\eeq
and the condition to observe interference reads
\andy{condint}
\beq
\triangle x \lesssim \triangle x_M \quad
\Leftrightarrow \quad \lambda_{L} \gtrsim 2d.
\label{eq:condint}
\eeq
In words, interference is preserved if the laser wavelength is
larger than twice the distance between the slits because in such a
case, by observing the scattered photon, one is unable to decide
which slit the photon came from.

Let us now consider a slightly different experiment [Figure
\ref{fig:2sllaser}(b)]. The laser spot is now focused at a
distance $\ell < L$ from the screen. From simple geometrical
considerations one gets
\andy{deltathetanew}
\beq
\triangle \theta \simeq \frac{\triangle p}{p} \simeq
\frac{h}{\lambda_{\rm L}p} \quad \Leftrightarrow \quad
\triangle' x \simeq \frac{h}{\lambda_{\rm L}p} \ell
\label{eq:deltathetanew}
\eeq
and the condition to observe interference reads now
\andy{condintnew}
\beq
\triangle' x \lesssim \triangle x_M \quad
\Leftrightarrow \quad \lambda_{L} \gtrsim 2d\frac{\ell}{L}.
\label{eq:condintnew}
\eeq
The physical reason is simple: if the laser spot is far from the
slits, the interfering waves converging to a given point of the
screen are closer to each other and one needs light of smaller
wavelength to resolve them. Notice that when $\ell \ll L$ the
laser wavelength needed to destroy the interference pattern
becomes very small!

The situation just described is somewhat reminiscent of Wheeler's
``delayed choice" \cite{Wheeler}, the important difference being
that in our case the longer the choice (to determine the route) is
postponed, the more ``effective" the measurement needs to be.

\section{Determining the trajectory by a laser beam}
\andy{compact}
\label{sec-compact}

So far the interfering system has been a structureless particle.
In the previous section we endeavored to obtain information on
the particle's route by scattering light on the system (notice
also that the scattering process was assumed to occur
instantaneously). However, we aim at describing a more complicated
physical picture, that can arise when the interfering system is
endowed with a richer internal physical structure. For instance,
consider the interference of C$_{60}$ molecules: in order to
obtain path information, one might shine laser light on the
molecule after it has gone through the slits, exactly like with
the Heisenberg-Bohm microscope. However, the situation would be
different, because the molecule can be regarded as a mesoscopic
system, whose inner structure is rich enough to give rise to more
complicated processes, involving lifetimes, emission of blackbody
radiation \cite{C60,Kolodney} and complex ionization processes
\cite{C60e,Hansen1,Hansen2,fullJMO}.
Unlike with the ``elementary" particle in the Heisenberg-Bohm
microscope, a fullerene molecule can absorb one or more photons
and undergo internal structural rearrangements.

It is therefore of great interest to try and understand how the
coherence properties of a ``mesoscopic" system are modified when
light of a given wavelength is shined on it, but the reemission
process takes place after a certain characteristic time. This
brings us conceptually closer to the situation envisaged in Figure
\ref{fig:2sllaser}(b). Needless to say, this is a simplified
picture of what would occur in the experiment performed by the
Vienna group \cite{AZfull} if one would try to obtain information
about the path of a fullerene molecule by illuminating it with an
intense laser beam. We shall come back to this point in the next
section, where a more realistic model will be considered. For the
moment, according to what we saw in the preceding section, the
minimal requirement to maintain quantum coherence and preserve the
interference pattern is the condition (\ref{eq:condintnew}).
However, we shall see that this is not the only criterion.

We start our considerations from a simple field-theoretical model.
This model is too elementary to reflect the complicated physical
effects that take place, for instance, in a fullerene molecule.
However, it has two main advantages: first, by virtue of its
simplicity, it admits an almost exact solution; second, in spite
of its simplicity, it captures some fundamental aspects related
to the notion of quantum mechanical coherence, when the
interfering system is more complicated than, say, an electron or
a neutron. Consider the Hamiltonian \cite{Cohen}
\andy{tothaml1,2,3}
\barr
 H & = & H_0 + V + V_{\rm L},
       \label{eq:tothaml1} \\
 H_0  & = & \frac{\bm p^2}{2m}+\hbar \omega_0 \ket{e}\bra{e} + \sum_i
\hbar \omega_i a^\dagger_{i} a_{i} , \label{eq:tothaml2} \\
 V & = & \sum_i
\left( \Phi_i e^{i\bmsub k_i\cdot\bmsub x} \ket{e}\bra{g} a_i
+\mbox{h.c.}
\right), \label{eq:tothaml3}\\
V_{\rm L}(t) & = & \left( \Phi_{\rm L}(t) e^{-i\omega_{\rm
L}t+i\bmsub k_{\rm L}\cdot\bmsub x}
\ket{e}\bra{g}
+\mbox{h.c.}
\right), \label{eq:tothaml4}
\earr
where
\andy{matrixel}
\barr
\Phi_i & = & -i e\bm d\cdot \bm\epsilon_i
\sqrt{\frac{\hbar\omega_i}{2\epsilon_0 L^3}},
\nonumber \\
\Phi_{\rm L}(t) & = & -i e\bm d\cdot \bm E_{\rm L} (t),
\label{eq:matrixel} \\
\bm d & = & \bra{e}\bm x\ket{g}.
\nonumber
\earr
We work in 3 dimensions. The above Hamiltonian describes a two-level
system (to be called ``molecule") of mass $m$, (center of mass)
position $\bm x$ and momentum $\bm p$, coupled to the electromagnetic
field, whose operators obey boson commutation relations
 \andy{boscomm2}
 \beq
 [a_i, a^\dagger_j] =\delta_{ij},
 \label{eq:boscomm2}
 \eeq
where the indexes $i,j$ are shorthand notations for the photon
momentum $\bm k_i$ and polarization $\lambda=1,2$. The ground
state $\ket{g}$ has energy 0, while the excited state $\ket{e}$
has energy $\hbar\omega_0$. The molecule interacts with a
(classical) laser, in the rotating-wave and dipole approximations.
The laser has electric field $\bm E_{\rm L}$ and frequency
$\omega_{\rm L}=c|\bm k_{\rm L}|$; we shall also assume that the
laser beam is parallel to the $y$-axis. The quantities $-e\bm d,
\bm\epsilon_i, \epsilon_0, L^3$ in (\ref{eq:matrixel}) are the
electric dipole moment, photon polarization, vacuum permittivity
and volume of the quantization box, respectively.

\begin{figure}
\begin{center}
\epsfig{file=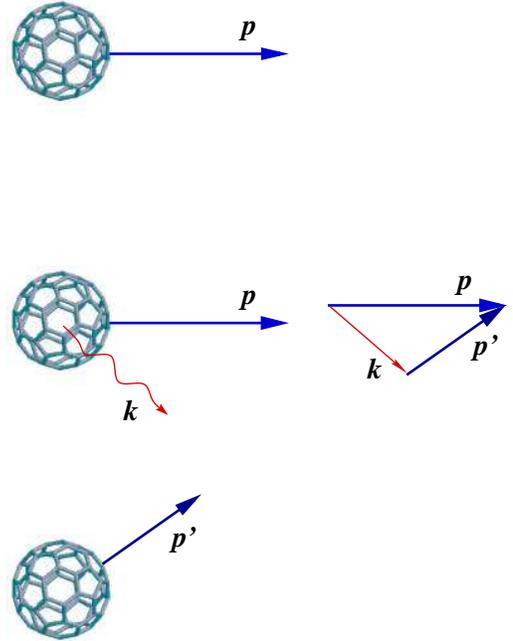,width=7cm}
\end{center}
\caption{Spontaneous emission process of a photon.}
\label{fig:nuovofull}
\end{figure}

The state of the total system will be written
\andy{statotot}
\beq
\ket{\Psi_{\rm
tot}}=\ket{\Psi,\alpha,n_i}\equiv\ket{\Psi}
\otimes\ket{\alpha}\otimes\ket{n_i},
\label{eq:statotot}
\eeq
where $\Psi$ denotes the spatial part of the wave function of the
molecule (in notation identical to that of Section 2),
$\alpha=e,g$ and $n_i$ is the number of photons emitted in the
$i$-mode during the $e$-$g$ transition. The state emerging from
the two slits is \andy{gausslrtot} \beq
\ket{\Psi_0,g,0}\equiv\ket{\Psi_0}\otimes\ket{g}\otimes\ket{0},
\label{eq:gausslrtot} \eeq where $\ket{\Psi_0}$ is given in
(\ref{eq:gausslr}). We assume that the laser beam is placed
immediately beyond the slits and illuminates coherently both wave
packets $|\Psi_\ell\rangle$ and $|\Psi_r\rangle$ in
(\ref{eq:gausslr}), like in Figure \ref{fig:2sllaser}(a). After a
laser pulse of duration $T$ such that $\int_0^T dt\;\Phi_{\rm
L}(t) e^{-i\omega_{\rm L} t}/\hbar=\pi/2$, the molecule has
absorbed a photon with probability 1 and the state reads
 \andy{gausslre}
\beq
\ket{\Psi_{\rm tot}}=\ket{e^{i\bmsub k_{\rm L}\cdot\bmsub
x}\Psi_0,e,0}.
\label{eq:gausslre}
\eeq
This is our ``initial" state. Since $\bm k_{\rm L}$ is parallel to
the $y$-axis, the molecule recoils along the vertical direction
without modifying the properties of the interference pattern (in
the $x$ direction). The spontaneous emission process is studied
in Appendix A, where the evolution is readily computed in the
Weisskopf-Wigner approximation \cite{seminal,seminal2,Merz} and
yields
\andy{WWsol}
\barr
\ket{\Psi_{\rm tot}(t)}&=&e^{-i\omega_0 t}e^{-\gamma t/2}
\ket{e^{-ip^2t/2m\hbar} e^{i\bmsub k_{\rm L}\cdot\bmsub x}\Psi_0,e,0}\nonumber\\
& & +\sum_i e^{-i\omega_i t} \beta_i(t) \nonumber\\
& & \times
\ket{e^{-ip^2t/2m\hbar} e^{i(\bmsub k_{\rm L}-\bmsub k_i)\cdot\bmsub x}\Psi_0,g,1_i},
\label{eq:WWsol}
\earr
where ($\alpha=e^2/4\pi\epsilon_0\hbar c$)
\andy{gammadef}
\barr
\gamma &=& \frac{2\pi}{\hbar^2}\sum_i|\Phi_i|^2 \delta(\omega_i-\omega_0)
\nonumber\\
&=& \frac{2\pi}{\hbar^2}\sum_\lambda\int
d^3k\;\frac{e^2\hbar\omega}{2\epsilon_0(2\pi)^3} |\bm
d\cdot\bm\epsilon_{\bmsub k \lambda}|^2
\delta(\omega-\omega_0) \nonumber\\
&=& \frac{4}{3}\frac{\alpha \omega_0^3 |\bm d|^2}{c^2},
\label{eq:gammadef}
\earr
is the decay rate (the third expression is the continuum limit),
as given by the Fermi ``golden" rule
\cite{Fermigold}, and
\andy{betaidef}
\beq
\beta_i(t)=\frac{\Phi^*_i}{\hbar} \frac{1-e^{i(\omega_i-\omega_0)t-\gamma t/2}}
{(\omega_i-\omega_0)+i\gamma/2} .
\label{eq:betaidef}
\eeq
The spontaneous emission process of a photon is shown in Figure
\ref{fig:nuovofull}. The total momentum is conserved and the
molecule recoils.
We see that in (\ref{eq:WWsol}) the internal degrees of freedom of
the molecule get entangled with the photon field, so that the
states in (\ref{eq:WWsol}) are all orthogonal each other. We can
now analyze the influence of the spontaneous emission process on
the interference pattern, i.e.\ on the quantum mechanical
coherence of the molecule. The intensity at the screen is readily
written as
\andy{intcorr}
\barr
I'(x)&=&\bra{\Psi_{\rm tot}(t_0)} x\rangle\bra{x}\Psi_{\rm
tot}(t_0)\rangle
\nonumber\\
&=&\exp(-\gamma t_0)\bra{\Psi^{\bmsub k_{\rm L}}(t_0)} x\rangle\bra{x}
\Psi^{\bmsub k_{\rm L}}(t_0)\rangle \nonumber\\
& & + \ \sum_i |\beta_i(t_0)|^2 \nonumber\\
& & \quad\quad \times
\bra{\Psi^{\bmsub k_{\rm L}-\bmsub k_i}(t_0)} x\rangle
 \bra{x}
\Psi^{\bmsub k_{\rm L}-\bmsub k_i}(t_0)\rangle \nonumber\\
&=&\exp(-\gamma t_0) I_{\bmsub k_{\rm L}}(x)+
\sum_i |\beta_i(t_0)|^2 I_{\bmsub k_{\rm L}-\bmsub k_i}(x),
\nonumber \\
\label{eq:intcorr}
\earr
where
\barr
\ket{\Psi^{\bmsub k}(t_0)}=
\exp\left(-i\frac{p^2}{2m\hbar}t_0\right)\exp(i\bm k\cdot \bm x)\ket{\Psi_0}
\earr
is the free evolution of the ``double Gaussian" wave packet that
has (jointly) recoiled (due to photon emission and/or absorption)
by momentum $\hbar\bm k$. In the position representation
\barr
\Psi^{\bmsub k}(\bm x,t_0) &=&
\bra{\bm x}\exp\left(-i\frac{p^2}{2m\hbar}t_0\right)
\exp(i\bm k\cdot \bm x)\ket{\Psi_0}
\nonumber \\
&=&
\frac{e^{-ip^2t_0/2m\hbar}
e^{i\bmsub k\cdot\bmsub x}}{\sqrt{N}}
[\Psi_\ell(\bm x)+\Psi_r(\bm x)].
\earr
The quantity $I_{\bmsub k_{\rm L}-{\bmsub k_i}}(x)$ represents the
partial interference pattern of those molecules that have emitted
a photon of momentum $\hbar\bm k_i$ (and absorbed a laser photon
of momentum $\hbar\bm k_{\rm L}$). By applying the same method
utilized for (\ref{eq:gaussscr})-(\ref{eq:intensity}), it is straightforward to obtain
\andy{intensityk}
\barr
I_{\bmsub k}(x)&=&\int dy dz|\Psi^{\bmsub k}(\bm x,t_0)|^2
\nonumber \\
&=&\frac{1}{N\sqrt{2\pi\delta x^2(t_0)}}
\left[
\exp\left(-\frac{\left(x-v_xt_0+\frac{d}{2}\right)^2}{2\delta x^2(t_0)}\right)
\right. \nonumber \\
& & \left. + \
\exp\left(-\frac{\left(x-v_xt_0-\frac{d}{2}\right)^2}{2\delta x^2(t_0)}\right)
\right. \nonumber \\ & &
\left.
+ \ 2
\exp\left(-\frac{(x-v_xt_0)^2+\left(\frac{d}{2}\right)^2}{2\delta
x^2(t_0)}\right)
\right.
\nonumber \\ & &
\left. \times \cos \left(\frac{\eta_x(t_0) d}{2\delta x^2(t_0)}
(x-v_x t_0)\right)
\right]\nonumber\\
&=& I\left(x-v_x t_0\right),
\label{eq:intensityk}
\earr
where $v_x=\hbar k_x/m$ is the $x$-component of the average
velocity (remember that $\bm k_0$ is parallel to $z$, so that
$v_x$ only gets a contribution from the emitted photon's $\bm k$).
Neglecting $a=d/2$ with respect to $\delta x(t_0)$ in the envelope
function we obtain
\andy{intensityk1}
\barr
I_{\bmsub k}(x) &\simeq&
\frac{e^{-(x-v_xt_0)^2/2\delta x^2(t_0)}}{\sqrt{2\pi\delta x^2(t_0)}}
\nonumber \\
& & \times \left[1+\cos\left(\frac{2\pi}{X}(x-v_x
t_0)\right)\right],
\label{eq:intensityk1}
\earr
where we set $X=h t_0/m d$, like in (\ref{eq:capX}). By recalling
that the laser beam is parallel to the $y$ direction, that is
$k_{{\rm L}x}=0$, and by noting that $I_{\bmsub k}(x)$ in
(\ref{eq:intensityk})-(\ref{eq:intensityk1}) depends only on
$k_x$, the intensity pattern (\ref{eq:intcorr}) reads
\andy{intcorrx}
\beq
I'(x)=\exp(-\gamma t_0) I(x)+
\sum_i |\beta_i(t_0)|^2 I_{-\bmsub k_i}(x).
\label{eq:intcorrx}
\eeq
As we can see, the interference pattern is made up of two terms: the
first one is associated with those molecules that have not emitted
any photon, the second one with those molecules that have emitted a
photon and recoiled accordingly. Obviously, the latter term depends
on the features of such emission.

Assume now that the spontaneous emission process is completely
isotropic, i.e.\ the directions of the dipole moments $\bm d$ in
(\ref{eq:matrixel}) of the molecules in the beam are completely
random. In this case the last term in Eq.\ (\ref{eq:intcorrx}) is
readily evaluated (see Appendix B) and yields
\andy{decayed}
\barr
& & \left\langle\sum_i |\beta_i(t_0)|^2 I_{-\bmsub
k_i}(x)\right\rangle
\nonumber \\
& & = (1-e^{-\gamma t_0}) \int \frac{d\Omega_{\bmsub {\bar
k}}}{4\pi}\; I_{-\bmsub {\bar k}}(x),
\label{eq:decayed}
\earr
where $|\bm {\bar k}|=\omega_0/c$ and $\langle\dots\rangle$
denotes the average over the molecular dipole direction. By using
(\ref{eq:intensityk1}), the last integral (average over the
direction of the emitted photon) yields
\andy{media}
\barr
& & \int d\Omega_{\bmsub {\bar k}}\; I_{-\bmsub {\bar k}}(x)
\nonumber \\
& & = 2\pi\int_{-1}^1 d\xi\;
\frac{e^{-(x+\bar v t_0 \xi)^2/2\delta x^2(t_0)}}{\sqrt{2\pi\delta x^2(t_0)}}
\nonumber \\
& & \quad \times
\left[1+\cos\left(\frac{2\pi}{X}(x+\bar v t_0\xi)\right)\right] ,
\label{eq:media}
\earr
where we set $\bar v_x/\bar v=\xi$. It is evident from this
expression that when $\bar v t_0=X/2$ the cosine is averaged over
the whole interval $2\pi$ and the second interference term in
(\ref{eq:intcorrx}) is completely washed out. For smaller values
of $\bar v t_0$ there is still some interference.

\begin{figure}
\begin{center}
\epsfig{file=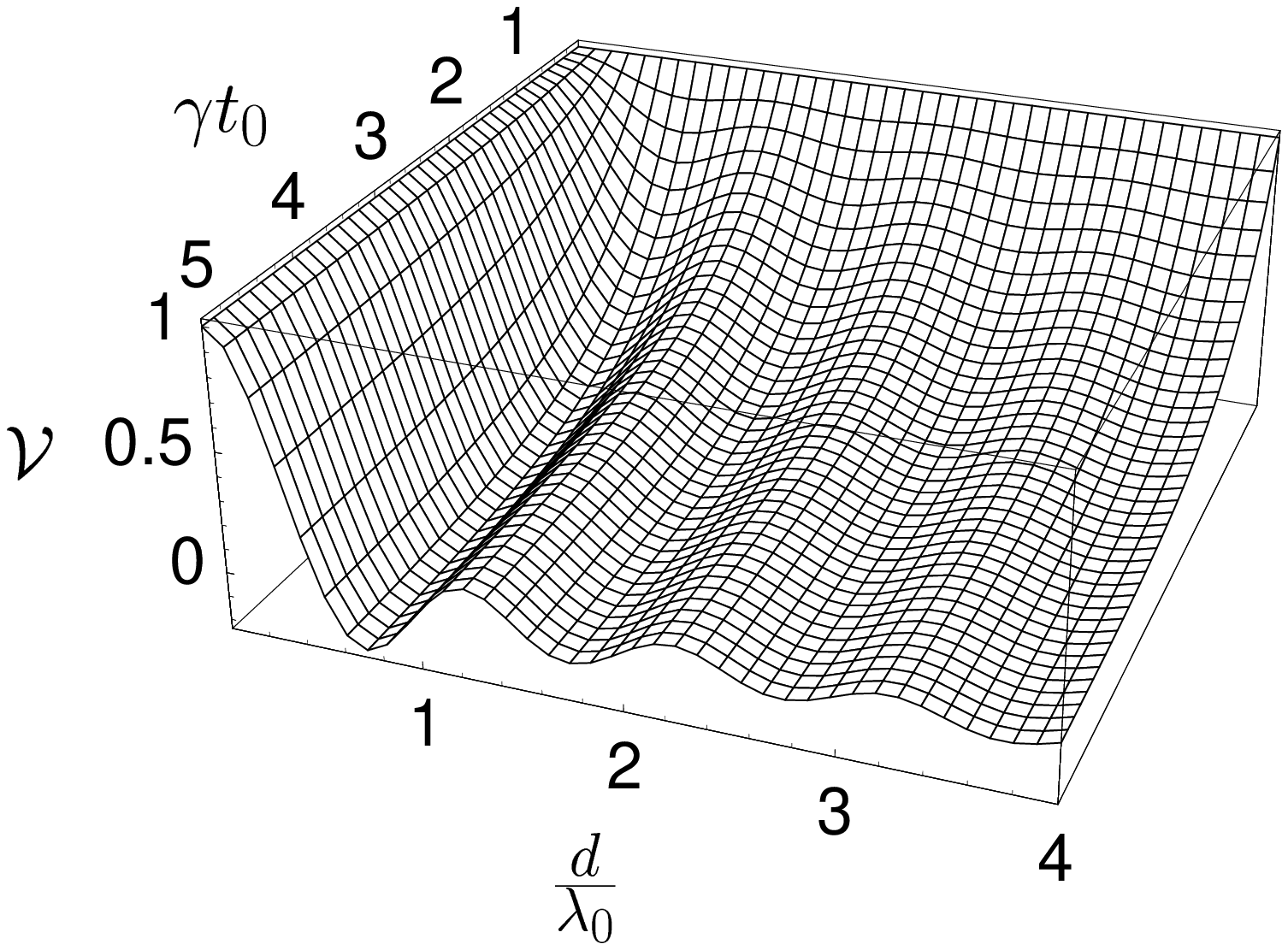,width=7cm}
\end{center}
\caption{The function ${\cal V}$ in Eq.\ (\ref{eq:visibility}).}
\label{fig:visgamlam}
\end{figure}

Let us focus on a realistic situation. Unlike in Section 2, we set
here $X/\delta x (t_0) = 0.4$ for clarity of presentation, in
order to get quite a few oscillations in the interference pattern
(this will also enable us to obtain compact expressions). In this
case $\bar v t_0<X/2\ll 2 \delta x(t_0)$ and the Gaussian envelope
in (\ref{eq:media}) is practically constant over the range of
integration. ($\exp(-X^2/8\delta x^2(t_0))=0.98$). We can then
write
\andy{media1}
\barr
& & \int d\Omega_{\bmsub {\bar k}}\; I_{-\bmsub {\bar k}}(x)
\simeq 2\pi\frac{e^{-x^2/2\delta x^2(t_0)}}{\sqrt{2\pi\delta
x^2(t_0)}}
\nonumber \\
 & & \quad \times
\int_{-1}^1 d\xi\;
\left[1+\cos\left(\frac{2\pi}{X}(x+\bar v t_0\xi)\right)\right]\nonumber\\
& & = 4\pi\frac{e^{-x^2/2\delta x^2(t_0)}}{\sqrt{2\pi\delta x^2(t_0)}}
\nonumber\\
& &
\quad \times \left[1+\sinc\left(\frac{\omega_0 d}{c}\right) \cos\left(\frac{2\pi}{X} x\right)\right]
,\label{eq:media1}
\earr
where $\sinc(x)\equiv \sin x/x$ and we used the equality $2\pi
\bar v t_0/X= \omega_0 d/c$. By plugging (\ref{eq:media1}) and
(\ref{eq:decayed}) into (\ref{eq:intcorrx}) we finally obtain
\andy{intfin}
\beq
I'(x)=\frac{e^{-x^2/2\delta x^2(t_0)}}{\sqrt{2\pi\delta x^2(t_0)}}
\left[1+{\cal V}\left(\gamma t_0, \frac{d}{\lambda_0}\right)
\cos\left(\frac{2\pi}{X} x\right)\right],
\label{eq:intfin}
\eeq
where
\andy{visibility}
\beq
{\cal V}\left(\gamma t_0, \frac{d}{\lambda_0}\right)= e^{-\gamma
t_0}+\left(1-e^{-\gamma t_0}\right)
\sinc\left(\frac{2\pi d}{\lambda_0}\right)
\label{eq:visibility}
\eeq
is shown in Figure \ref{fig:visgamlam} as a function of
$d/\lambda_0$ and $\gamma t_0$.

\begin{figure}
\begin{center}
\epsfig{file=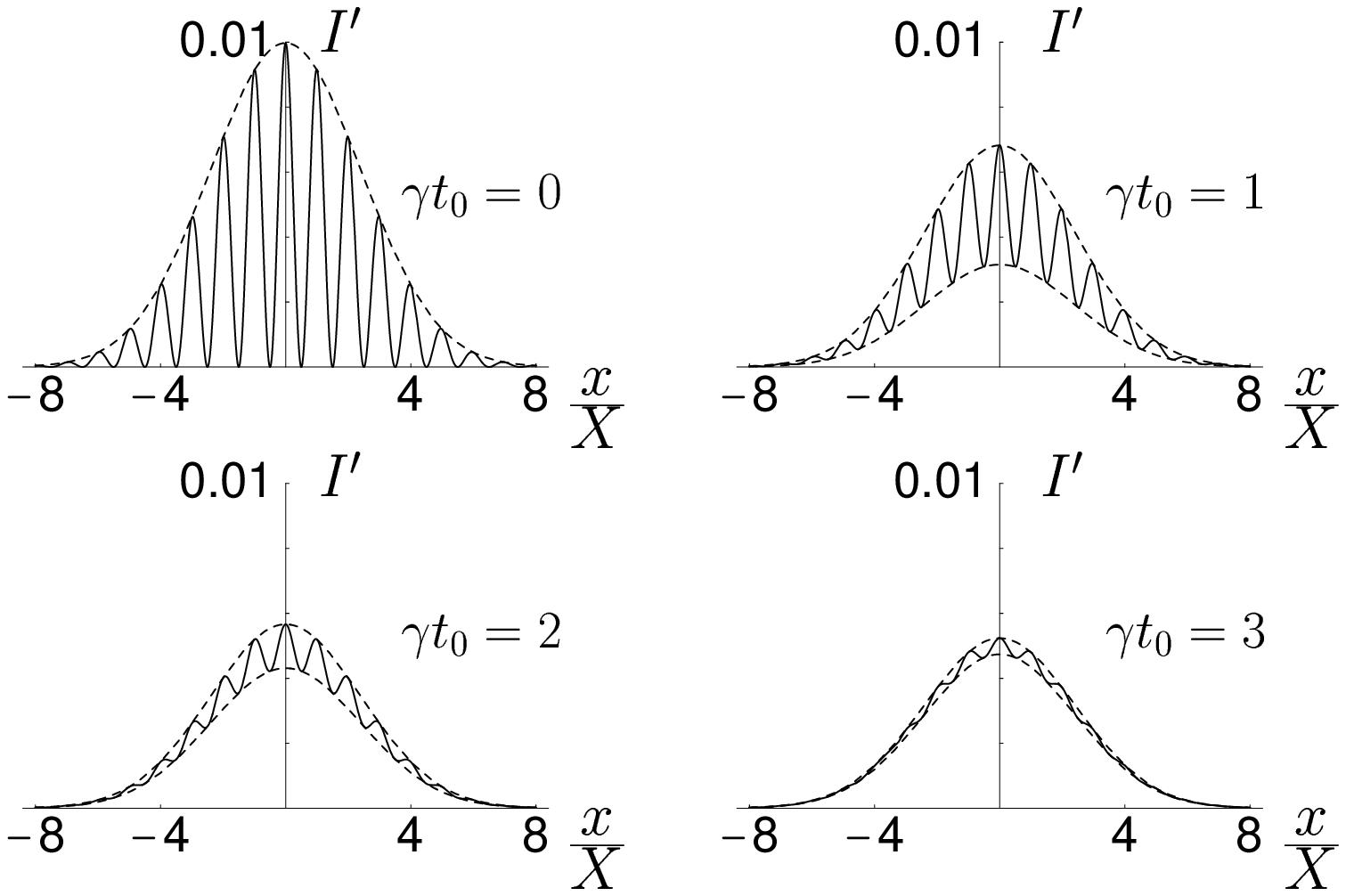,width=8cm}
\end{center}
\caption{Intensity at the screen when $d/\lambda_0=2$,
$X/\delta x(t_0)=0.4$ and $\gamma t_0$ is varied. From top left to
bottom right, $\gamma t_0=0,1,2,3$. }
\label{fig:fringdeco}
\end{figure}

\begin{figure}
\begin{center}
\epsfig{file=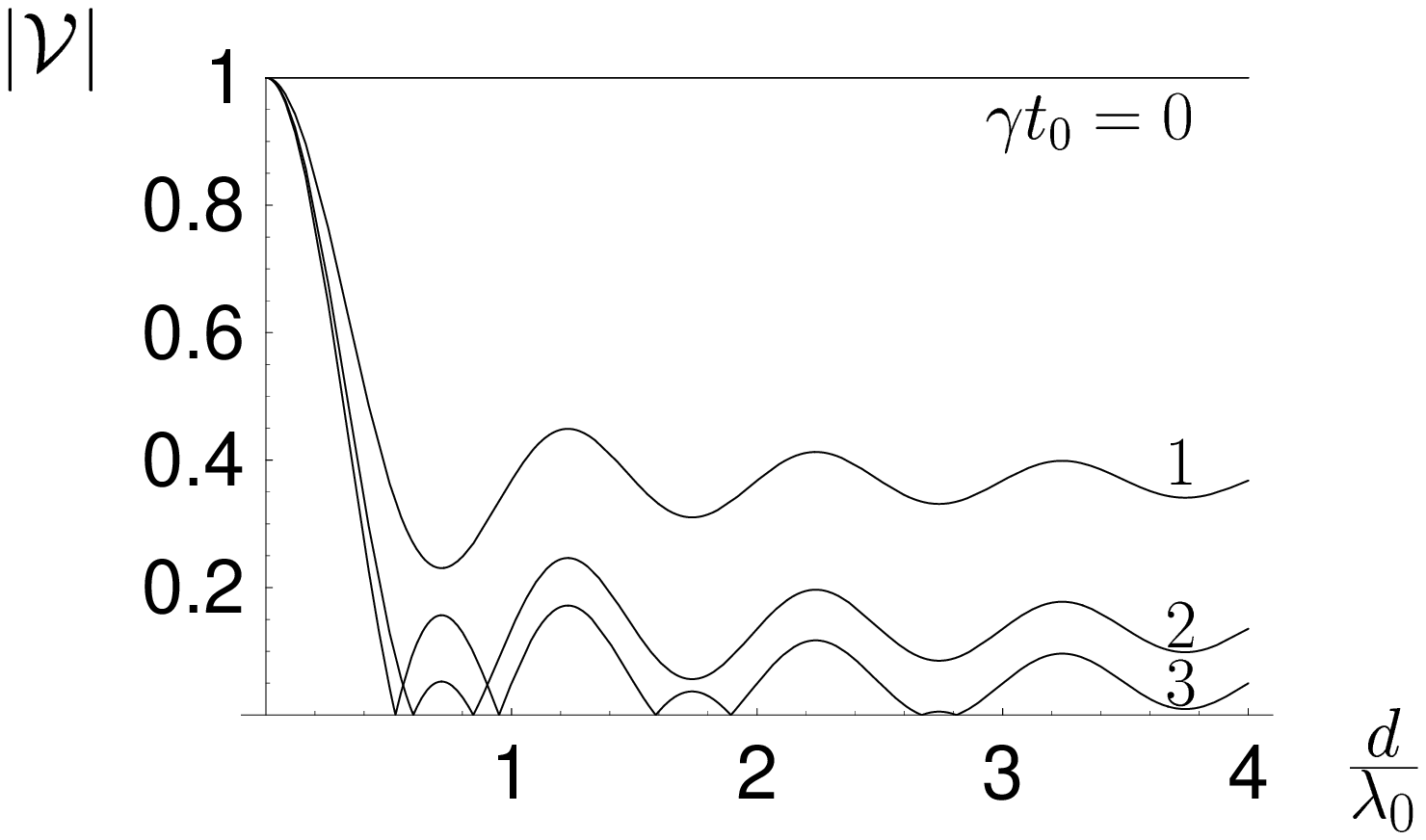,width=8.5cm}
\end{center}
\caption{The visibility of the interference pattern as a function
of $d/\lambda_0$, for $\gamma t_0=0,1,2,3$.}
\label{fig:vislambda}
\end{figure}

We now look at some particular cases. The intensity at the screen
is displayed in Figure \ref{fig:fringdeco} for $d/\lambda_0=2$ and
a few values of $\gamma t_0$. The quantity $|{\cal V}|$ is the
visibility of the interference pattern
\andy{standvis}
\beq
|{\cal V}|=\frac{I'_{\rm max}-I'_{\rm min}}{I'_{\rm max}+I'_{\rm
min}}\ .
\label{eq:standvis}
\eeq
 Roughly speaking, the visibility is related to the amplitude of the oscillations of
the interference pattern and measures the degree of coherence of
the interfering system \cite{fibonacci}. Notice that the
visibility decreases as $\gamma t_0$ is increased, namely when the
emission process of the photon is faster. This is readily
understood in terms of the discussion in Section 3 ($\lambda_0$
plays the role of $\lambda_{\rm L}$). The behavior of the
visibility as a function of $d/\lambda_0$ is shown in Figure
\ref{fig:vislambda} for the same values of $\gamma t_0$ as those
used in Figure \ref{fig:fringdeco}.

In order to appreciate the meaning of these results, let us first
observe that the interpretation of the visibility derives from
(\ref{eq:intcorrx}): the first term in the r.h.s.\ of
(\ref{eq:visibility}) is associated with those molecules that have
not emitted any photon (and reach the screen in an excited state),
while the second term is associated with those molecules that have
emitted a photon before they hit the screen.
\noindent
When the wavelength of the photon satisfies the coherence
condition (\ref{eq:condint}), $\lambda_0 \gtrsim 2d$, by detecting
the emitted photon we cannot extract any path information and the
visibility reads
\beq
{\cal V}\left(\gamma t_0, \frac{d}{\lambda_0} \lesssim \frac{1}{2}
\right)\simeq {\cal V}\left(\gamma t_0, 0\right)=1 :
\eeq
the interference pattern is equal to that obtained when no laser
is present [namely, (\ref{eq:intfin}) reduces to
(\ref{eq:freepatt})], irrespectively of the value of $\gamma t_0$.

Let now $\lambda_0<d$, so that the coherence condition
(\ref{eq:condint}) is {\em not} satisfied. We clearly see from
Figure \ref{fig:vislambda} that, somewhat unexpectedly, coherence
is {\em still largely} preserved if $\gamma t_0 \lesssim 1$,
because even though the photon wavelength is small enough to yield
information about the path of the interfering particle, such a
path information is {\em not accessible}: it is, so to say,
``stored" in the internal structure of the molecule. Such an
information would be available to an external observer {\em only}
if the photon were emitted. Mathematically,
\beq
{\cal V}\left(\gamma t_0, \frac{d}{\lambda_0}\gtrsim 1
\right)\simeq {\cal V}\left(\gamma t_0, \infty\right)=\exp(-\gamma
t_0)
\eeq
which tends to vanish if the decay is rapid ($\gamma t_0\gg 1$)
and to unity if the decay is slow ($\gamma t_0\ll 1$).

In conclusion, the interference pattern is blurred out (${\cal
V}\simeq0$), only if the photon emission process yields {\it both}
a good resolution, $\lambda_0\lesssim d$, {\it and} a quick
response, $\gamma t_0\gg1$. We recover in this case the
conclusions of the Heisenberg-Bohm microscope analyzed in Section
3: if the decay is rapid we have the situation shown in Figure
\ref{fig:2sllaser}(a), while if the decay is slower we are closer to the
case depicted in Figure \ref{fig:2sllaser}(b) [yielding the less
stringent condition (\ref{eq:condintnew})]. Formally, the
Heisenberg-Bohm microscope of Figure
\ref{fig:2sllaser}(a) is fully recovered in the (familiar) limit
\andy{visibilitysinc}
\beq
{\cal V}\left(\infty, \frac{d}{\lambda_0}
\right)=
\sinc\left(\frac{2\pi d}{\lambda_0}\right) .
\label{eq:visibilitysinc}
\eeq
It is interesting to notice that space and time considerations are
both important in this context: in order to lose quantum
coherence, the molecule must interact with its environment in such
a way that its path information is not only available, but also
{\em quickly} available. This is a significant difference with the
Heisenberg-Bohm microscope: a good ``resolution" is needed, both
in space and time.

\begin{figure}
\begin{center}
\epsfig{file=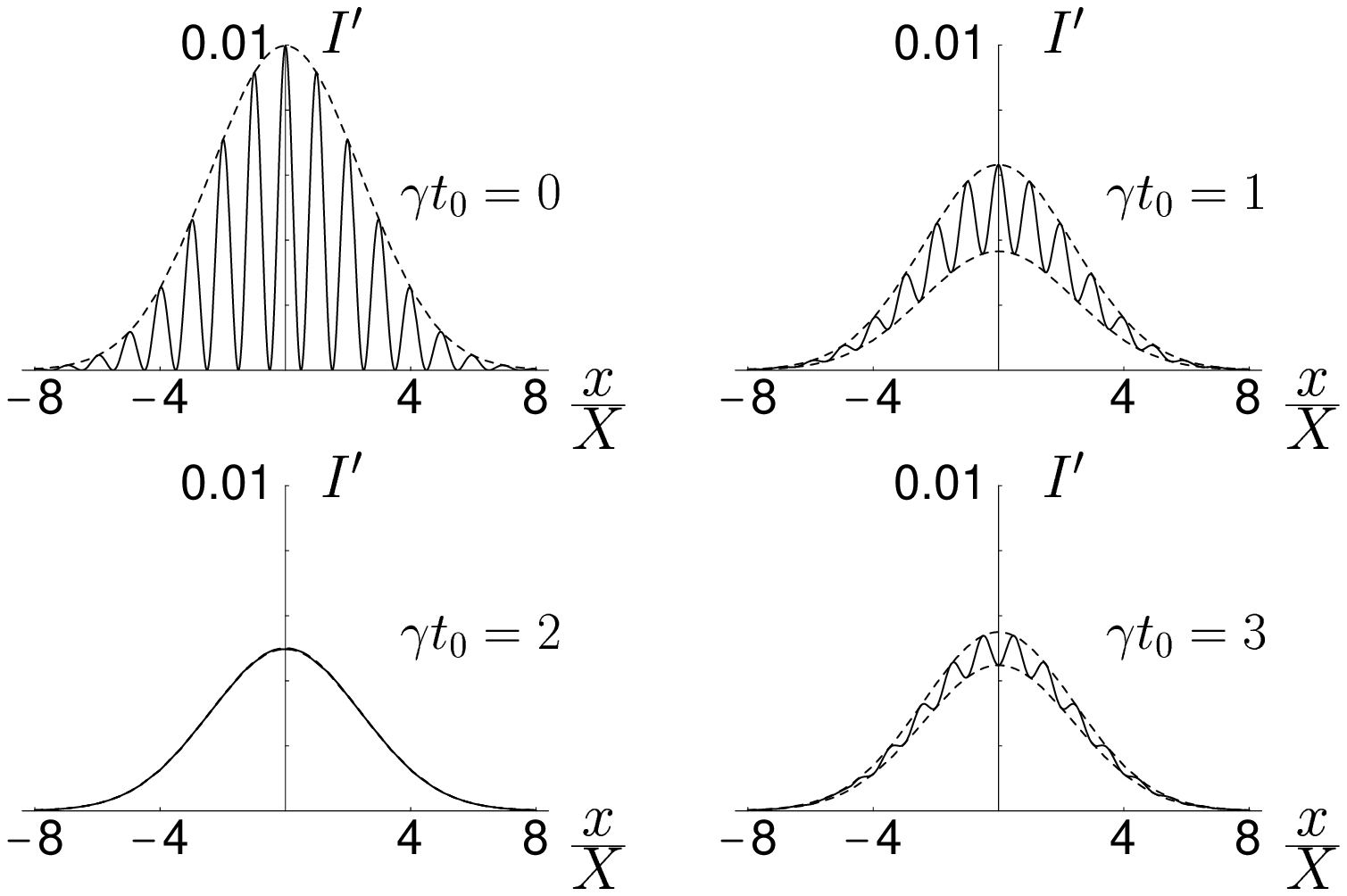,width=8cm}
\end{center}
\caption{Intensity at the screen when $d/\lambda_0=0.84$,
$X/\delta x(t_0)=0.4$ and $\gamma t_0$ is varied. From top left to
bottom right, $\gamma t_0=0,1,2,3$. }
\label{fig:fringinvers}
\end{figure}

There is more. One might be led to think that the visibility (and
therefore the quantum coherence) is always a decreasing function
of $d/ \lambda_0$: in other words, a smaller photon wavelength
(yielding better path information) always increases decoherence.
This expectation is incorrect: look at Figure \ref{fig:vislambda},
where the visibility exhibits in general an oscillatory behavior.
``Regular" regions, where the visibility decreases by decreasing
the  wavelength, are interspersed with ``anomalous" regions in
which by decreasing the photon wavelength the visibility
increases: a better microscope does not necessarily yield more
information. One infers that there are physical situations in
which the behavior of the visibility is somewhat ``anomalous" and
at variance with naive expectation. Similar cases were
investigated in neutron optics \cite{fibonacci} and are related to
well-known phenomena in classical optics (see for instance Sec.\
7.5.8 of Ref.\ \cite{BornWolf}).

\begin{figure}
\begin{center}
\epsfig{file=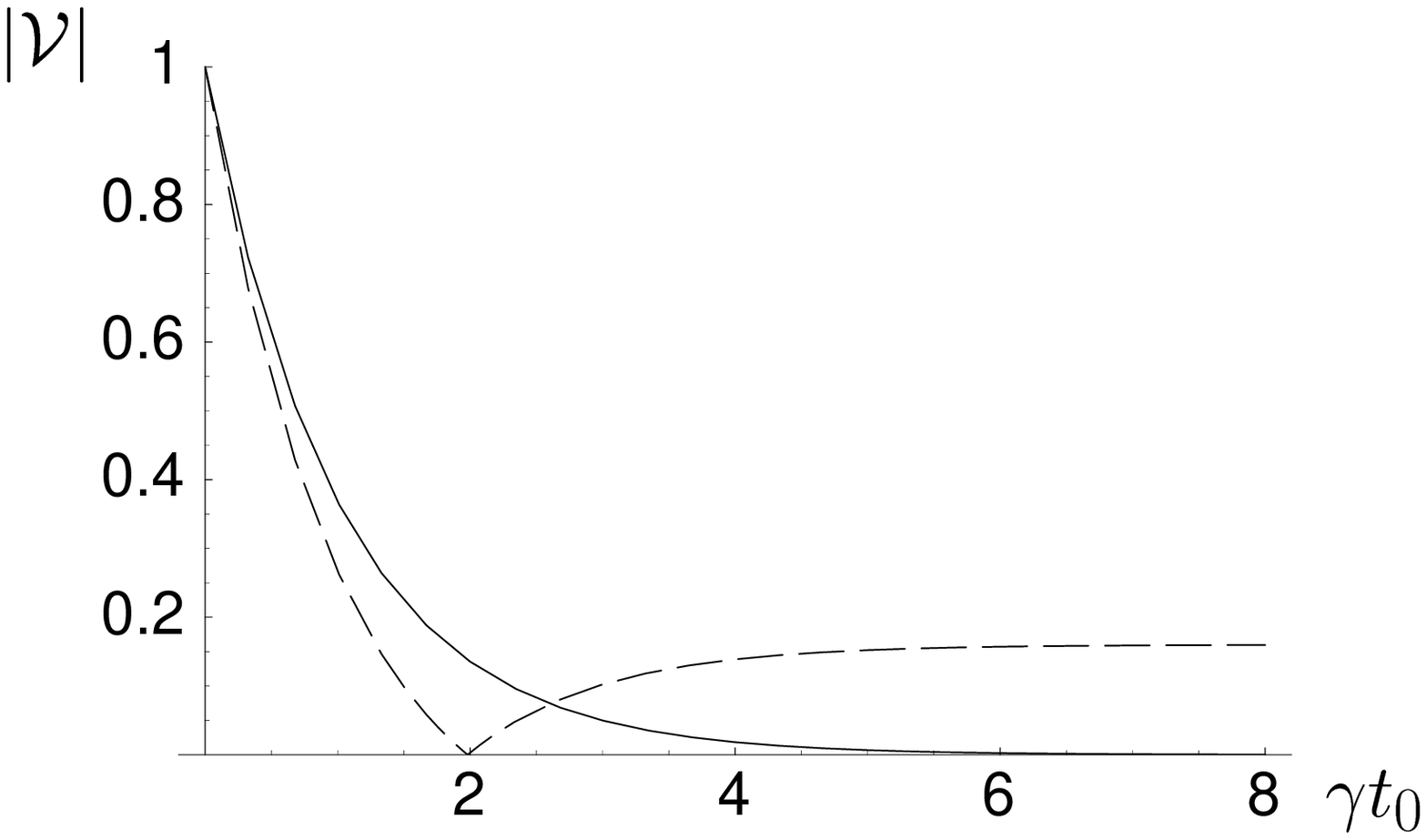,width=8cm}
\end{center}
\caption{The visibility of the interference pattern as a function
of $\gamma t_0$, for $d/\lambda_0=0.84$ (dashed line) and $2$
(continuous line). The former situation is ``anomalous."}
\label{fig:visgamma}
\end{figure}

The above discussion deals with spatial resolution. A similar
phenomenon occurs also in time domain, where a faster photon
emission (yielding path information) does not necessarily increase
decoherence. In Figure \ref{fig:fringinvers} the intensity at the
screen is shown for $d/\lambda_0=0.84$ and a few values of $\gamma
t_0$. The visibility reaches a minimum (in fact vanishes) for
$\gamma t_0=2$ and then increases again. (This phenomenon appears
together with an interchange of minima and maxima.) Notice the
difference with Figure \ref{fig:fringdeco}. The behavior of the
visibility for the cases shown in Figures
\ref{fig:fringdeco} and
\ref{fig:fringinvers} is displayed in Figure \ref{fig:visgamma} as
a function of $\gamma t_0$.

\begin{figure}
\begin{center}
\epsfig{file=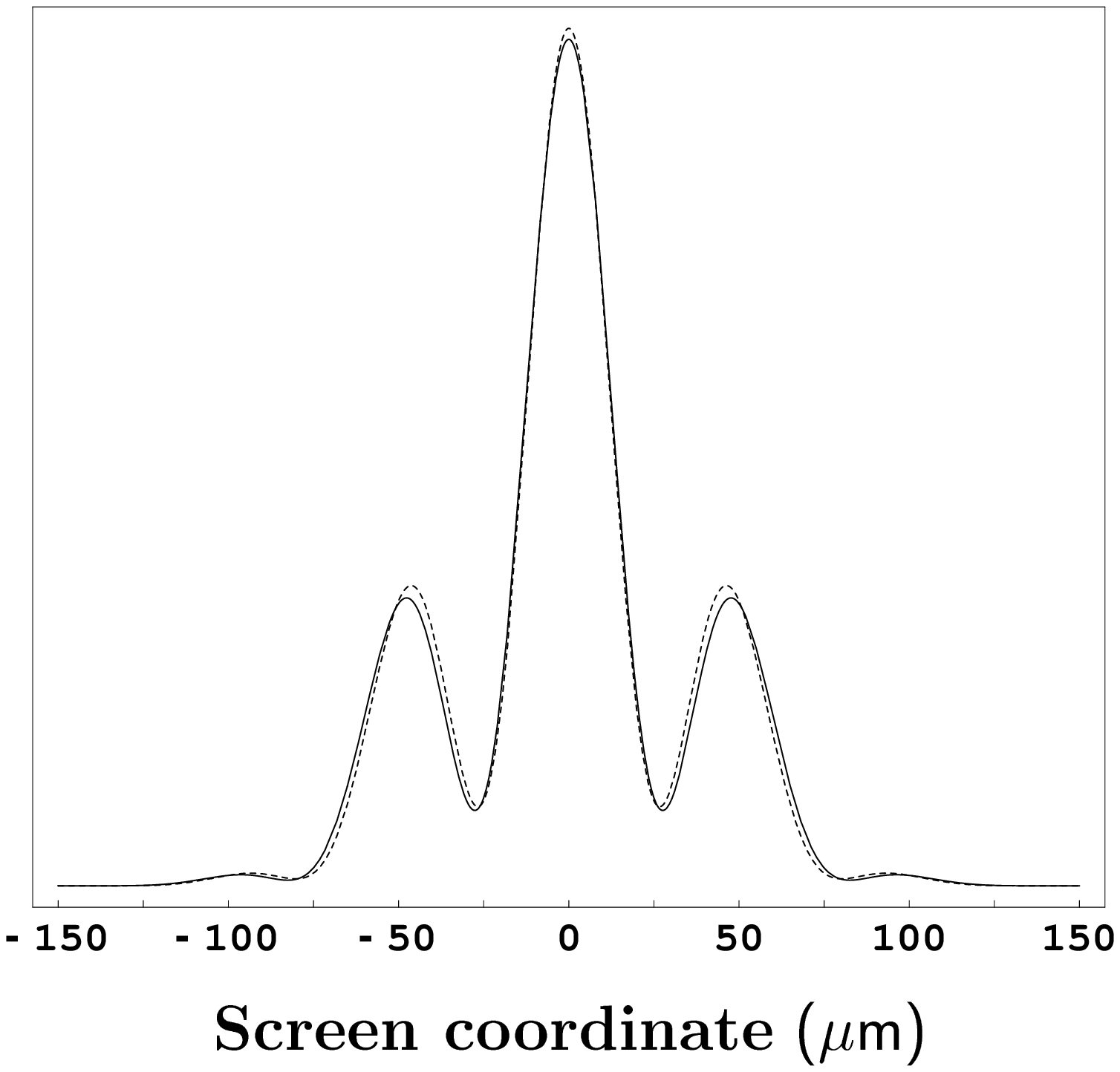,width=6.45cm}
\end{center}
\caption{Interference patterns for $X/\delta x(t_0)=1.56$.
The full line is the exact formula, obtained by
(\ref{eq:intcorrx})-(\ref{eq:media}), the dashed line the
approximate expression (\ref{eq:intfin}). Compare with Figure
\ref{fig:pattern}.}
\label{fig:intcompare}
\end{figure}

We conclude with an additional comment. For the numerical values
considered in Section 2, one gets $X/\delta x(t_0)=1.56$ and
$\exp(-X^2/8\delta x^2(t_0))=0.74$, so that there are only a few
oscillations within the envelope function, as can be seen in
Figures \ref{fig:pattern} and \ref{fig:exptresult}. However, the
assumptions leading to the expression (\ref{eq:visibility}) for
the visibility [see the paragraph preceding (\ref{eq:media1})]
maintain their validity. The interference patterns given by the
approximate expression (\ref{eq:intfin}) and by the exact formulas
(\ref{eq:intcorrx})-(\ref{eq:media}) are shown in Figure
\ref{fig:intcompare}: they are almost identical.

\section{A more refined picture of the molecule}
\label{sec-nlevels}
\andy{nlevels}

A molecule of fullerene is a complicated object, that can absorb
several visible photons at once and undergo quite involved
processes in its internal structure. The physical model analyzed
in the preceding section is too simple to describe such a rich
physical picture. Although it yields nice insight, the model is
unsatisfactory because it is not able to describe the absorption
and reemission process of several photons. This is what one would
need, because the physics of fullerene is for certain aspects
related to that of a small black body, characterized by a
well-defined temperature and in continuous interaction with its
environment \cite{fullJMO,Hansen2,Hansen1,Kolodney}.

It is possible to analyze the interference of fullerene by
introducing a more realistic (and complicated) model: a detailed
calculation is still feasible, but requires more sophisticated
techniques and will be presented elsewhere. However, we will
briefly outline some of the nice qualitative features of the
physical picture that emerges from such an analysis.

The mesoscopic system (molecule of fullerene) can still be
described by a model similar to the one introduced in the
preceding section: the molecule is viewed as a multi-level system
that starts its evolution, immediately after the slits, in a
highly excited state (or possibly in a mixed state of given {\em
temperature}). On its way to the screen, the molecule emits some
(say $N =n \pm \triangle n$) photons of different energies and in
random directions. Of course, unlike in the previous section, the
photons have low energy, although the sum of their energies can be
significant (and for instance comparable with the energy of the
single photon emitted in the preceding section).

\begin{figure}
\begin{center}
\epsfig{file=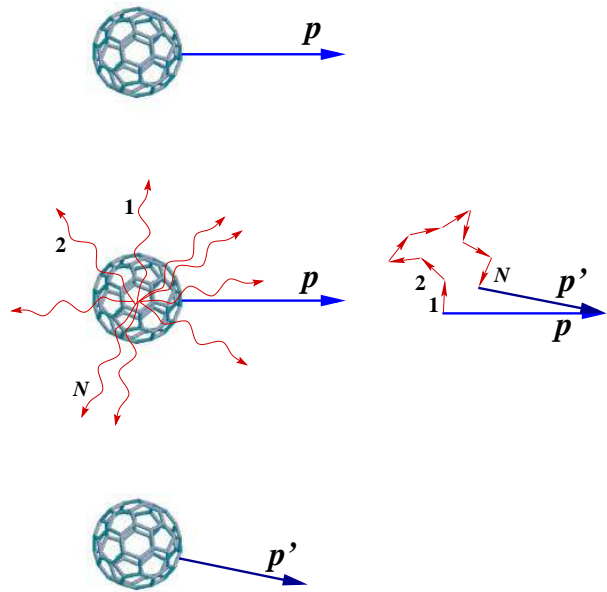,width=8cm}
\end{center}
\caption{Random walk (\ref{eq:randomrec}) in momentum space.
$N$ (thermal) photons are emitted and the momentum of the molecule
changes accordingly. }
\label{fig:randomwalk}
\end{figure}

\begin{figure}
\begin{center}
\epsfig{file=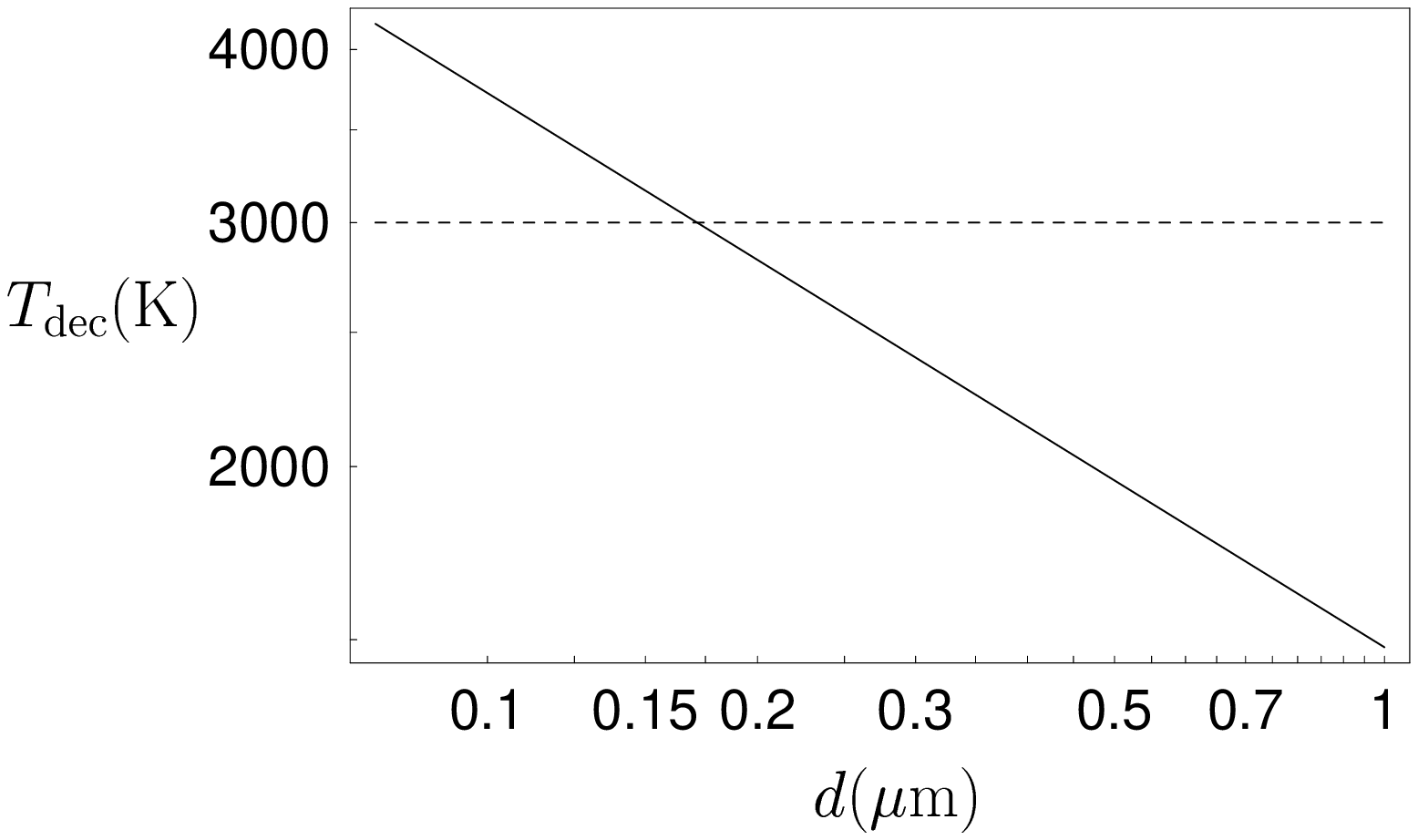, width=8.5cm}
\end{center}
\caption{$T_{\rm dec}$ versus $d$ (slit separation) for a fullerene molecule.
We set the time of flight $t_0=9.53$ms. The horizontal line is the
temperature at which fragmentation becomes significant. }
\label{fig:tvslnd}
\end{figure}

The question is: will the interference pattern be modified as a
consequence of the multiple emission processes that take place
between the slits and the screen? The answer is: less than one
might think. In order to justify this statement at a
semiquantitative level, look at Figure \ref{fig:randomwalk}. The
photons will be emitted in random directions and as a consequence
the momentum of the molecule will recoil by the quantity
\andy{randomrec}
\beq
\triangle p \simeq \hbar\langle k \rangle \sqrt{n},
\label{eq:randomrec}
\eeq
where $\hbar\langle k \rangle$ is the average momentum of the
emitted photons and $n$ the average number of emitted photons. The
molecule, as a consequence of light emission processes, loses a
total energy $\triangle E \simeq n \hbar\langle k
\rangle/c$ between the slits and the screen. However, according to
(\ref{eq:randomrec}), its momentum will
only be changed by the quantity
\andy{Erec}
\beq
\triangle p \simeq \frac{\triangle E}{c\sqrt{n}}.
\label{eq:Erec}
\eeq
For instance, the interference pattern will be only slightly
affected by the emission of a large number of low-energy photons.
This is an interesting qualitative conclusion.
\noindent
A more quantitative relation can be obtained by treating fullerene
as a macroscopic system that emits thermal radiation at
temperature $T$. The total intensity of emission $J_0$ and the
total photon flux $\Phi_0$ emitted by a black body read
\cite{Landau}
\andy{emiss,flux}
\barr
J_0&=&\frac{c E}{4V}=\frac{\hbar}{4\pi^2c^2}\int_0^\infty
d\omega\;\frac{\omega^3}{e^{\hbar\omega/k_{\rm B}
T}-1}\nonumber\\& &=\frac{\pi^2}{60}\frac{k_{\rm
B}^4}{c^2\hbar^3}T^4 ,
\label{eq:emiss}\\
\Phi_0&=&\frac{c N_V}{4V}=\frac{1}{4\pi^2c^2}\int_0^\infty
d\omega\;\frac{\omega^2}{e^{\hbar\omega/k_{\rm B} T}-1}\nonumber\\
& & =\frac{\zeta(3)}{2\pi^2}\frac{k_{\rm B}^3}{c^2\hbar^3}T^3 ,
\label{eq:flux}
\earr
respectively, where $k_{\rm B}$ is the Boltzmann constant, $E$
($N_V$) the total radiation energy (number of photons) contained
in a cavity of volume $V$ and
$\zeta(s)=\sum_{n=1}^\infty\frac{1}{n^s}$ the Riemann function
($\zeta(3)\simeq1.202$). The energy and number of photons emitted
by the surface $A$ of the fullerene molecule during its time of
flight $t_0$ are, respectively,
\andy{DEn}
\beq
\triangle E=J_0 A t_0, \qquad n=\Phi_0 A t_0.
\label{eq:DEn}
\eeq
Therefore by making use of (\ref{eq:DEn}), (\ref{eq:emiss}) and
(\ref{eq:flux}), Eq.\ (\ref{eq:Erec}) reads
\andy{ErecT}
\barr
\triangle p &\simeq& \frac{J_0 A t_0}{c\sqrt{\Phi_0 A t_0}}
=\kappa (A t_0)^{\frac{1}{2}} T^{\frac{5}{2}}, \label{eq:ErecT1} \\
\kappa&=&\sqrt{\frac{2}{\zeta(3)}}\frac{\pi^3}{60}\frac{k_{\rm
B}^{\frac{5}{2}}}{c^2 \hbar^{\frac{3}{2}}}\nonumber\\&=&
4.85\times10^{-24}\mbox{ kg / s$^{\frac{3}{2}}$
K$^{\frac{5}{2}}$}.
\label{eq:ErecT2}
\earr
In agreement with Section 3, in order to observe interference, the
transferred momentum $\triangle p$ must satisfy the inequality
(coherence condition, obtained by (\ref{eq:condint}) and
(\ref{eq:deltap}))
\andy{conservo}
\beq
\triangle p \lesssim \frac{h}{2 d} ,
\label{eq:conservo}
\eeq
which translates into the following bound for the internal
temperature $T$
\barr
T &\lesssim& \xi \frac{1}{(A t_0
d^2)^{\frac{1}{5}}} \equiv T'_{\rm dec} \label{eq:TT} \\
\xi & = &
\left(\frac{1800\;\zeta(3)}{\pi^4}\right)^{\frac{1}{5}}
 \frac{\hbar\; c^{\frac{4}{5}}}{k_{\rm B}} \nonumber \\
&=& 8.59\times10^{-5}\mbox{ K s$^{\frac{1}{5}}$ m$^{\frac{4}{5}}$}.
\earr
By taking $A=4\pi r^2=1.539\times10^{-18}$m ($r\simeq 3.5$\AA\ is
the radius of a fullerene molecule) and $t_0=9.47$ms
\cite{AZcomptes} we get
\beq
T'_{\rm dec}\simeq 500 \mbox{K} .
\eeq
This bound is too strong: a fullerene molecule cannot be
considered as an ordinary black body. Its curvature cannot be
neglected \cite{C60e,Kolodney,Hansen2} and its emitting surface is
far from being flat. One can take a heuristic approach and
summarize its behavior by multiplying the quantities in Eq.\
(\ref{eq:DEn}) by an emissivity coefficient $\alpha
\simeq 4.5\cdot10^{-5}$ (due to the curvature of the emitting surface
for small atomic clusters \cite{Kolodney}), to obtain
\andy{DEna}
\beq
\triangle E= \alpha J_0 A t_0, \qquad n= \alpha \Phi_0 A t_0,
\label{eq:DEna}
\eeq
so that
\andy{ErecT2a}
\beq
\triangle p \simeq \sqrt{\alpha}\kappa (A t_0)^{\frac{1}{2}} T^{\frac{5}{2}},
\label{eq:ErecT2a}
\eeq
This yields
\andy{boundcoh}
\beq
T \lesssim \xi
\alpha^{-\frac{1}{5}}\frac{1}{(A t_0d^2)^{\frac{1}{5}}}
\equiv T_{\rm dec} .
\label{eq:boundcoh}
\eeq
This is a more reliable estimate, that should be valid at least as
an order of magnitude. Equation (\ref{eq:boundcoh}) is a {\em
coherence condition}. The quantity $T_{\rm dec}$ is the internal
(``blackbody") temperature of a fullerene molecule at which
decoherent effects should become apparent in a double slit
experiment. For the numerical values of the Vienna experiment
\cite{AZcomptes,fullJMO}
\beq
T_{\rm dec} \simeq 3700 {\rm K} .
\eeq
Notice that above $T\simeq 3000$K fullerene molecules begin to
fragmentate (ionization is likely to occur at even lower
temperatures). We are led to argue that the temperature of the
fullerene molecule will only have a small influence on the
visibility of the interference pattern, at least for the Vienna
experimental configuration. However, if the experiment is modified
by letting the fullerene go through an interferometer of the
Mach-Zender type in order to increase the beam separation $d$
(say, up to a distance of order $1\mu$m), then intrinsic
decoherence effects should come to light. The behavior of $T_{\rm
dec}$ versus $d$ (slit separation) is shown in Figure
\ref{fig:tvslnd} for a time of flight of 9.53ms (distance
travelled $L=1.22$m and speed $v_z=128$m/s \cite{AZcomptes}).
Decoherence effects should be visible at about 2000K for a beam
separation of order of half a micron. (Notice: in such a
situation, according to (\ref{eq:DEna}), the molecule emits
$n\simeq 8.6$ photons.)
\noindent
We stress that the calculation of this section is based on a {\em
heuristic} model and is probably valid within a numerical factor
of order unity. Indeed we have not considered a few effects that
should yield interesting corrections: first, we have neglected the
temperature of the environment (assuming that it is much lower
than that of the molecule) and all cooling mechanisms of fullerene
(which is a {\em small} black body and loses energy by emitting
photons: since C$_{60}$ has 174 vibrational modes, 4 of which are
infrared active, the temperature decrease during a time of flight
of 10ms should be of a few hundreds Kelvin). Second, we have
neglected all entanglement effects due to photon emissions into
the environment (such entanglement effects were automatically
taken into account in the calculation of the preceding section).
Third, it should be emphasized that our calculation is valid for
those fullerene molecules that reach the screen (detection system)
in an electrically neutral state: the very hot C$_{60}$ molecules
that yield C$_{60}^+$ ions {\em during} the flight to the screen
will strongly couple to environmental stray fields and will not
interfere. Since the beautiful detection mechanism in the Vienna
experiment hinges upon ionization, these additional ions should be
removed from the beam. Fourth, fragmentation effects have not been
considered. Finally, we notice that our estimate is based on the
(conservative) approximation (\ref{eq:conservo}).

From a merely theoretical viewpoint, it is interesting to observe
how one obtains sensible results in the mesoscopic domain by
combining thermodynamical considerations with a pure quantum
mechanical analysis.

\section{What is decoherence?}
\label{sec-conc}
\andy{conc}

Decoherence is an interesting phenomenon. Although the quantum
mechanical {\em coherence} is readily defined and is intuitively
related to the possibility of creating a superposition in a
Hilbert space, it is not obvious what the {\em lack} of quantum
coherence is. Most physicist would say that such a lack of
coherence can be given a meaning only in a statistical sense (but
there are noteworthy historical exceptions \cite{NPN}).

However, without endeavoring to give rigorous definitions, one has
the right to ask {\em when, how} and {\em why} coherence is
maintained. This is not an easy question, in particular if the
system investigated is not strictly microscopic. In this paper we
have studied a molecule with an increasingly complicated internal
structure. If the molecule is ``elementary," i.e.\ structureless,
the usual description in terms of the Heisenberg-Bohm microscope
applies and interference is lost when a photon of suitable
wavelength is scattered off the molecule after the latter has gone
through a double slit. If, on the other hand, the molecule can be
reasonably schematized as a two-level system, the situation is not
that simple. One still needs a photon of suitable wavelength to
destroy interference, but in addition the photon reemission
process must be rapid. If, for instance, the photon is reemitted
only after the molecule has reached the screen, no {\em Welcher
Weg} information is available and interference (coherence) is
preserved. In words, the two-level system needs a certain time to
``explore" its environment, e.g.\ via photon emission, and ``give
away" {\em Welchew Weg} information. Note: via photon emission,
{\em not} photon absorption!

If the picture becomes more complicated and the molecule can
absorb and reemit a large number of photons, then the situation
becomes even more interesting. In such a case the molecule might
be complicated enough to have an internal {\em temperature} and
one can properly talk of ``mesoscopic interference." In such a
situation, the mesoscopic system will slowly ``explore" its
environment by emitting photons in the course of its evolution and
its branch waves will slowly ``decohere" (namely, their momentum
will recoil due to repeated photon emissions and their states
propagating through different slits will get entangled with
increasingly orthogonal states of the electromagnetic field, that
plays the role of environment). In this sense, coherence (viewed
as ability to interfere) simply means isolation from the
environment---a lesson that experimental physicists know much
better (and since much longer) than their theoretical colleagues,
who do not (and never had to) care about ``isolating" a beam in
order to make it interfere.

We notice that there is no big conceptual change if the molecule
reemits an electron, rather than a photon, after a certain
characteristic (life)time due to internal rearrangements. For
example, ions C$_{60}^+$ are quickly produced after multiphoton
excitation with laser pulses, for wavelengths ranging from
ultraviolet (200 nm) to infrared (1000 nm) light. (It is just this
emission process of an electron with well-defined energy
\cite{C60e,Hansen1} that was brilliantly used to detect the
fullerene molecules in \cite{AZfull,fullJMO}.) Electron emission
gives rise to nonnegligible recoil and therefore the interference
pattern will vanish upon averaging over the emission direction.
More so, coupling to stray electric field would prevent any
reasonable isolation of C$_{60}^+$ from the environment.

We are left with an interesting question, though. When can a
system be described by a wave function? When is a quantum state
``pure"? The experiment \cite{AZfull} has taught us something
fundamental. If a good experimentalist can {\em isolate} the
internal degrees of freedom of a (microscopic, mesoscopic or even
{\em macro}scopic) system and at the same time succeeds in
controlling the coherent superposition of an additional dynamical
variable (say, the coordinate of the center of mass of the
molecule) then the state
$$
\mbox{superposed wave function $\times$ internal state}
$$
will interfere (we apologize for the abuse of notation and
additional inaccuracies). Notice that the internal state can be
mixed and can even be a {\em black body}. If there were a party
(with a few cats) going on inside the fullerene molecule and if
the external world would not be able to notice, {\em not even in
principle}, the molecule would still interfere. Flabbergasting.

\acknowledgments
We thank M.\ Arndt, G.\ van der Zouw and A.\
Zeilinger for useful discussions and for kindly giving us detailed
information about their fullerene interference experiment. Figure
\ref{fig:exptresult} is reproduced with their permission. We also
acknowledge stimulating remarks by A.\ Scardicchio. This work was
done within the framework of the TMR European Network on ``Perfect
Crystal Neutron Optics" (ERB-FMRX-CT96-0057).

\renewcommand{\thesection}{\Alph{section}}
\setcounter{section}{0}
\setcounter{equation}{0}
\section{The Weisskopf-Wigner approximation and the Fermi ``golden" rule}
\label{sec-appA}
\andy{appA}

\renewcommand{\thesection}{\Alph{section}}
\renewcommand{\thesubsection}{{\it\Alph{section}.\arabic{subsection}}}
\renewcommand{\theequation}{\thesection.\arabic{equation}}

\begin{figure}
\begin{center}
\epsfig{file=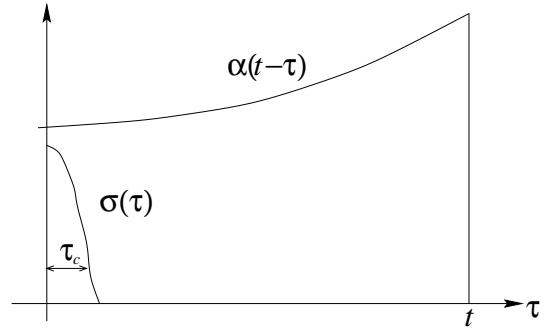, width=7cm}
\end{center}
\caption{The functions $\alpha$ and $\sigma$.}
\label{fig:brown}
\end{figure}

We study here the evolution of the system introduced in Section 4
and derive Eqs.\ (\ref{eq:WWsol})-(\ref{eq:betaidef}). The initial
state is (\ref{eq:gausslre}) and the evolution reads
\andy{evolint}
\beq
\ket{\Psi_{\rm tot}(t)}=e^{-iH_0 t/\hbar} U_{I}(t)\ket{\Psi_{\rm tot}}
=e^{-iH_0 t/\hbar} \ket{\Psi_{\rm tot}(t)}_I,
\label{eq:evolint}
\eeq
with
\andy{Uint}
\barr
U_I(t) &=& T\exp\left(-\frac{i}{\hbar}\int_0^t
d\tau\;V_I(\tau)\right),\label{eq:Uint} \\
V_I(\tau) &=& e^{iH_0\tau/\hbar} V e^{-iH_0\tau/\hbar},
\earr
where $I$ denotes the interaction picture and $T$ time ordering.
We study the emission process by solving the Schr\"odinger
equation in the interaction picture
\andy{Schrint}
\beq
\ket{\dot \Psi_{\rm tot}(t)}_I=-\frac{i}{\hbar} V_I(t)
\ket{\Psi_{\rm tot}(t)}_I.
\label{eq:Schrint}
\eeq
In the notation of Section 4, the interaction Hamiltonian is
\andy{interinter}
\barr
V_I(t)\!&=&\! \sum_i \left[ \Phi_i e^{-i
\left(\omega_i-\frac{\hbar k_i^2}{2m}-\omega_0\right)t} e^{i
\bmsub k_i \cdot\bmsub x} e^{i \frac{\bmsub k_i\cdot \bmsub
p}{m}\; t} \ket{e}\bra{g} a_i
\right.\nonumber\\
& &\quad\quad  +\mbox{h.c.}\Bigg]
\label{eq:interinter}
\earr
and by writing the state of the system as
\andy{stateint}
\barr
\ket{\Psi_{\rm tot}(t)}_I&=&\alpha(t)
\ket{e^{i\bmsub k_{\rm L}\cdot\bmsub x}\Psi_0,e,0}\nonumber\\
& & +\sum_i \beta_i(t) e^{i \left(\delta_i-\frac{\hbar k_i^2}{2m}\right) t} \nonumber\\
& & \times
\ket{e^{-i \frac{\bmsub k_i\cdot \bmsub p}{m}\; t} e^{i(\bmsub k_{\rm L}-\bmsub k_i)\cdot\bmsub x}\Psi_0,g,1_i},
\qquad
\label{eq:stateint}
\earr
with
\andy{deltaphase}
\beq
\delta_i= \frac{\hbar\bm k_i}{m}\cdot(\bm k_0+
\bm k_{\rm L})-\frac{\hbar k_i^2}{2m},
\label{eq:deltaphase}
\eeq
the Schr\"odinger equation (\ref{eq:Schrint}) reads
\andy{coeffint}
\barr
\dot \alpha &=& -\frac{i}{\hbar}\sum_i \beta_i(t) \Phi_i
e^{-i(\bar \omega_i-\omega_0)t},
\nonumber\\
\dot \beta_i&=&
 -\frac{i}{\hbar}\alpha(t) \Phi_i^* e^{i(\bar
\omega_i-\omega_0)t} ,
\label{eq:coeffint}
\earr
where $\bar\omega_i=\omega_i-\delta_i$. Incorporating the initial
conditions $\alpha(0)=1$, $\beta_i(0)=0$, Eqs.\
(\ref{eq:coeffint}) are formally solved to yield
\andy{coeffint2}
\barr
\dot \alpha(t) &=& - \sum_i \frac{|\Phi_i|^2}{\hbar^2}
 \int_0^t d\tau\; e^{-i(\bar\omega_i-\omega_0)\tau}\alpha(t-\tau)
 .\nonumber\\
 \beta_i(t) &=& -\frac{i}{\hbar} \Phi_i^* \int_0^t d\tau\;
e^{i(\bar\omega_i-\omega_0)\tau}\alpha(\tau),
\label{eq:coeffint2}
\earr
The former is an integro-differential equation for $\alpha(t)$ and
can be rewritten in the form
\andy{integrodiff}
\beq
\dot \alpha(t) = -
 \int_0^t d\tau\; \sigma(\tau) \alpha(t-\tau),
\label{eq:integrodiff}
\eeq
where $\sigma(t)$ is the Fourier transform of the form factor
$\Gamma(\omega)$
\andy{Fourtrans,Gamma}
\barr
\sigma(t)&=&\int \frac{d\omega}{2\pi}\; \Gamma(\omega)
e^{-i(\omega-\omega_0)t},\label{eq:Fourtrans}\\
\Gamma(\omega)&=& \frac{2\pi}{\hbar^2} \sum_i |\Phi_i|^2 \delta(\bar\omega_i-\omega).
\label{eq:Gamma}
\earr
If $\Gamma(\omega)$ is a smooth function with bandwidth $\Lambda$,
by the uncertainty relation $\triangle t
\triangle\omega\simeq 1$, $\sigma(t)$ is significantly different
from zero only within a time interval $\tau_c\simeq 1/\Lambda$.
Therefore, the integral in the r.h.s.\ of Eq.\
(\ref{eq:integrodiff}) takes a substantial contribution only from
the time domain $(0,\tau_c)$, as shown in Figure \ref{fig:brown}.
If we assume that the evolution of $\alpha(t)$ takes place on a
time scale that is much longer than $\tau_c$, we get
$\alpha(t-\tau)\simeq\alpha(t)$ for $\tau\lesssim\tau_c$, whence
\andy{alphadot}
\beq
\dot \alpha(t) \simeq - \alpha(t)
 \int_0^t d\tau\; \sigma(\tau).
  \label{eq:alphadot}
\eeq
At times $t\gg\tau_c$, the integral becomes almost independent of
$t$, the integration can be extended up to $t=+\infty$ and we
obtain from (\ref{eq:Fourtrans})
\beq
 \int_0^t d\tau\; \sigma(\tau)\simeq\int_0^\infty d\tau \int \frac{d\omega}{2\pi}\; \Gamma(\omega)
e^{-i(\omega-\omega_0-i0^+)\tau},
\eeq
where we introduced an infinitely small negative imaginary term
$-i0^+$ in order to assure the convergence. By interchanging the
order of integration, we get
\barr
& &\int \frac{d\omega}{2\pi}\;\Gamma(\omega) \int_0^\infty d\tau
e^{-i(\omega-\omega_0-i0^+)\tau}\nonumber\\
& &=\int
\frac{d\omega}{2\pi}\;\Gamma(\omega)\frac{-i}{\omega-\omega_0-i0^+},
\earr
which, by making use of the well-known formula
\andy{Plemelij}
 \beq
 \lim_{\epsilon\rightarrow 0^+} \frac{1}{x-i\epsilon}=
 {\cal P}\frac{1}{x}
+ i\pi\delta(x),
 \label{eq:Plemelij}
 \eeq
where ${\cal P}$ denotes the principal part, becomes
\barr
& &\int_0^t d\tau\; \sigma(\tau)\simeq i\Delta+\frac{\gamma}{2},
\label{eq:Deltagamma}
 \\
\Delta&=&{\cal P}\int \frac{d\omega}{2\pi}\;
\frac{\Gamma(\omega)}{\omega_0-\omega}={\cal P}\sum_i
\frac{|\Phi_i|^2/\hbar^2}{\omega_0-\bar\omega_i},\;\;\;\;
\label{eq:Delta}
\\
\gamma&=&\Gamma(\omega_0)=\frac{2\pi}{\hbar^2} \sum_i |\Phi_i|^2
\delta(\bar\omega_i-\omega_0).
\label{eq:gamma}
\earr
The last two quantities are the second-order correction to the
energy $\hbar\omega_0$ and the Fermi ``golden" rule
\cite{Fermigold}, Eq.\ (\ref{eq:gammadef}) in the text. By
plugging Eq.\ (\ref{eq:Deltagamma}) into Eq.\ (\ref{eq:alphadot})
we finally get
\andy{WWappr}
\beq
\dot \alpha(t) = -\left(i \Delta + \frac{\gamma}{2}\right)
\alpha(t),
\label{eq:WWappr}
\eeq
which is the celebrated Weisskopf-Wigner approximation
\cite{seminal,seminal2}.
It simply consists in replacing a complex dynamics with memory
effects [Eq.\ (\ref{eq:integrodiff})] with a simpler one
characterized by a purely Markovian equation\ (\ref{eq:WWappr}),
giving rise to exponential decay
\andy{exponential}
\beq
\alpha(t) = \exp\left(-i \Delta t - \frac{\gamma}{2} t \right).
\label{eq:exponential}
\eeq
As we stressed before, this replacement is justified for
$t\gg\tau_c$ only if there exist two well separated time scales,
i.e.\ only if the amplitude probability $\alpha(t)$ of the excited
level $\ket{e}$ evolves much more slowly than the characteristic
time $\tau_c$ of the photon reservoir. By making use of Eq.\
(\ref{eq:exponential}) and the definitions
(\ref{eq:Delta})-(\ref{eq:gamma}) this condition translates into
the following inequality
\andy{condition}
\beq
\Delta^{-1}, \gamma^{-1} \gg \tau_c \quad\Leftrightarrow\quad
\Gamma(\omega_0)\ll \Lambda ,
\label{eq:condition}
\eeq
which is always verified for sufficiently small coupling and/or
large bandwidth $\Lambda$, such as QED decay processes in vacuum
[remember that $\Gamma(\omega)\propto |\Phi|^2\propto\;$(coupling
constant)$^2$].

By substituting the solution (\ref{eq:exponential}) in the second
equation (\ref{eq:coeffint2}) and performing the integration, one
gets
\andy{betaexp}
\beq
\beta_i(t) = \frac{\Phi^*_i}{\hbar}\frac{1-e^{i(\bar\omega_i-\omega_0- \Delta) t -
\gamma t/ 2}}{ \bar\omega_i-\omega_0- \Delta+i \gamma/ 2} ,
\label{eq:betaexp}
\eeq
which is Eq.\ (\ref{eq:betaidef}) of the text. By plugging
(\ref{eq:exponential}) and (\ref{eq:betaexp}) into
(\ref{eq:stateint}) and using (\ref{eq:evolint}) one gets Eq.\
(\ref{eq:WWsol}) of the text. In our analysis we will always
neglect the energy shift $\hbar \Delta$ or, equivalently, absorb
it into the (renormalized) free energy $\hbar
\omega_0$. We will also neglect the phase shift $\delta_i$, which
is due to recoil and Doppler effects. Indeed both terms in
(\ref{eq:deltaphase}) are much smaller than the transition
frequency $\omega_0$: $(\hbar k^2_i/2 m)/\omega_0\sim
\hbar\omega_0/2mc^2\ll 1$ and $(\hbar \bm k\cdot\bm k_0/m)/\omega_0\sim
v_z/c\ll 1$. Moreover, note that when we write the evolved state
$\ket{\Psi_{\rm tot}(t)}_I$ as in Eq.\ (\ref{eq:stateint}) we are
neglecting any contributions out of the relevant Tamm-Duncoff
sector, due to frequency-dependent Doppler and/or recoil effects,
which deform the Gaussian wave packet.

\setcounter{equation}{0}
\setcounter{figure}{0}
\section{Evaluation of an integral}
\label{sec-appB}
\andy{appB}

We want to derive Eq.\ (\ref{eq:decayed}). We start by computing
the integral
\andy{esercizio}
\beq
\sum_i |\beta_i(t)|^2 =
\sum_i \frac{|\Phi_i|^2}{\hbar^2} \; \frac{|1-e^{i(\omega_i-\omega_0)t-\gamma t/2}|^2}
{(\omega_i-\omega_0)^2+\gamma^2/4} .
\label{eq:esercizio}
\eeq
This is an interesting integral (often left as an exercise in good
textbooks \cite{Merz}), that must be performed, like other
calculations related to the Fermi ``golden" rule, with the help of
physical intuition. Expanding the square we find
\andy{espandi}
\barr
& &
\sum_i \frac{|\Phi_i|^2}{\hbar^2} \; \frac{1 + e^{-\gamma t}-2 e^{-\gamma t/2} \cos
(\omega_i-\omega_0)t}
{(\omega_i-\omega_0)^2+\gamma^2/4} \nonumber \\
& & =
(1+e^{-\gamma t}) \sum_i
\frac{|\Phi_i|^2/\hbar^2}{(\omega_i-\omega_0)^2+\gamma^2/4}
\nonumber \\
& & \quad - 2 e^{-\gamma t/2} \sum_i \frac{|\Phi_i|^2}{\hbar^2}\frac{\cos
(\omega_i-\omega_0)t}{(\omega_i-\omega_0)^2+\gamma^2/4}
\nonumber \\
& & = 1 + e^{-\gamma t} -2 e^{-\gamma t/2} \sum_i
\frac{|\Phi_i|^2}{\hbar^2}\frac{\cos (\omega_i-\omega_0)t }
{(\omega_i-\omega_0)^2+\gamma^2/4},\nonumber\\
\label{eq:espandi}
\earr
where the Fermi ``golden" rule (\ref{eq:gamma})
\andy{fgr}
\beq
\gamma= \frac{2\pi}{\hbar^2} \sum_i |\Phi_i|^2 \delta
(\omega_i-\omega_0)
\label{eq:fgr}
\eeq
has been used. Moreover, we obtain
\andy{intcon}
\beq
\int dx \frac{\cos xt} {x^2+\gamma^2/4} =
\int dx \frac{e^{ixt}} {x^2+\gamma^2/4} \nonumber \\
 = \frac{2\pi}{\gamma}e^{-\gamma t/2} ,
\label{eq:intcont}
\eeq
so that, in the spirit of the golden rule,
\andy{asint}
\beq
\frac{\cos xt} {x^2+\gamma^2/4} \stackrel{\gamma \to 0}{\sim}
\frac{2\pi}{\gamma}e^{-\gamma t/2}\delta (x).
\label{eq:asint}
\eeq
Plugging this result into Eq.\ (\ref{eq:espandi}) we get
\andy{fineeser}
\beq
\sum_i |\beta_i(t)|^2 =
1- e^{-\gamma t}.
\label{eq:fineeser}
\eeq
We can now tackle Eq.\ (\ref{eq:decayed}). This appears in the
form
\andy{newesercizio}
\barr
& & \sum_i |\beta_i(t)|^2 f_i \nonumber \\
& & =
(1+e^{-\gamma t}) \frac{2\pi}{\gamma}
\sum_i \frac{|\Phi_i|^2}{\hbar^2}f_i \delta (\omega_i-\omega_0)
\nonumber \\
& & \quad - 2 e^{-\gamma t/2} \frac{2\pi}{\gamma} e^{-\gamma t/2}
\sum_i \frac{|\Phi_i|^2}{\hbar^2}f_i\delta (\omega_i-\omega_0)
\nonumber \\
& & = (1-e^{-\gamma t}) \frac{2\pi}{\gamma}
\sum_i \frac{|\Phi_i|^2}{\hbar^2}f_i \delta (\omega_i-\omega_0)
.\quad
\label{eq:newesercizio}
\earr
The above derivation applies under the same conditions of Appendix
A, namely when $\gamma \ll$ bandwidth. Now consider the average of
Eq.\ (\ref{eq:newesercizio}) over the direction of the molecular
dipole $\bm d$
\andy{mediad}
\barr
& &\left\langle \sum_i |\beta_i(t)|^2 f_i\right\rangle
=\sum_i\left\langle |\beta_i(t)|^2
\right\rangle f_i\nonumber\\
& & \;\;=(1-e^{-\gamma t}) \frac{2\pi}{\gamma}
\sum_i \frac{\left\langle|\Phi_i|^2\right\rangle}{\hbar^2}f_i \delta
(\omega_i-\omega_0).\qquad
\label{eq:mediad}
\earr
Note that
\andy{Phimedio}
\barr
\left\langle|\Phi_i|^2\right\rangle&=&\frac{e^2 \hbar\omega_i}{2\epsilon_0
L^3}\frac{1}{4\pi}\int d\Omega_{\bmsub d} |\bm
d\cdot\bm\epsilon_i|^2
\nonumber\\
&=&\frac{e^2 \hbar\omega_i}{2\epsilon_0 L^3}\frac{1}{3} |\bm d|^2
\label{eq:Phimedio}
\earr
is only a function of $\omega_i$ and does not depend on the
direction of the photon momentum $\bm k_i$, whence
\andy{mediad1}
\barr
& &\left\langle \sum_i |\beta_i(t)|^2 f_i\right\rangle
\nonumber\\
& & \;\;=(1-e^{-\gamma t}) \frac{2\pi}{\gamma}\frac{e^2
\omega_0|\bm d|^2}{6\hbar\epsilon_0 L^3}
\sum_i f_i \delta
(\omega_i-\omega_0)\nonumber\\
& & \;\;=(1-e^{-\gamma t}) \frac{\pi^2 c^3}{\omega_0^2 L^3}
\sum_i f_i \delta
(\omega_i-\omega_0) .\qquad
\label{eq:mediad1}
\earr
In the continuum limit, if $f_i$ does not depend on the photon
polarization $\lambda$, we finally get
\andy{mediad2}
\barr
& &\left\langle \sum_i |\beta_i(t)|^2 f_i\right\rangle
\nonumber\\
& & \;\;=(1-e^{-\gamma t}) \frac{c^3}{\omega_0^2 4\pi}
\int d^3 k\; f_{\bmsub k} \delta
(\omega-\omega_0)\nonumber\\
& & \;\;=(1-e^{-\gamma t}) \frac{1}{4\pi}
\int d\Omega_{\bmsub {\bar k}}\; f_{\bmsub {\bar k}},
 \qquad
\label{eq:mediad2}
\earr
with $|\bm {\bar k}|=\omega_0/c$, which yields the result we
sought.


\end{document}